\documentclass[abstract=true, DIV=14]{scrartcl}

\usepackage[noEnd,commentColor=black]{algpseudocodex}
\usepackage{amsmath}
\usepackage{amssymb}
\usepackage{amsthm}
\usepackage[mathscr]{euscript}
\usepackage[shortlabels]{enumitem}
\usepackage{graphicx} 
\usepackage{subcaption}
\usepackage{thmtools}
\usepackage{todonotes}
\usepackage{xcolor}
\usepackage{microtype}

\usepackage[normalem]{ulem}

\usepackage{tcolorbox}

\usepackage{upgreek}
\usepackage[colorlinks]{hyperref}
\usepackage[capitalize]{cleveref}

\crefname{equation}{eq.}{eqs.} 
\crefname{enumi}{}{} 
\crefname{icase}{case}{cases}
\crefname{ipart}{part}{parts}
\crefname{iprop}{property}{properties}
\crefname{iinv}{invariant}{invariants}

\crefformat{section}{\S\,#2#1#3}
\crefformat{subsection}{\S\,#2#1#3}
\crefrangeformat{section}{\S\S\,#3#1#4--#5#2#6}
\crefmultiformat{section}{\S\S\,#2#1#3}{,~#2#1#3}{,~#2#1#3}{,~#2#1#3}

\usepackage{authblk}
\usepackage[normalem]{ulem}

\usepackage{tikz}
\usetikzlibrary{calc}

\newcommand{\fO}{\mathcal{O}}

\newcommand{\ttrue}{\mathtt{true}}
\newcommand{\tfalse}{\mathtt{false}}

\DeclareMathOperator{\ex}{ex} 

\DeclareMathOperator{\tw}{tw} 
\DeclareMathOperator{\gw}{gw} 

\newcommand{\sread}{\mathsf{read}}
\newcommand{\srank}{\mathsf{rank}}
\newcommand{\sselect}{\mathsf{select}}

\renewcommand{\alpha}{\upalpha}

\newcommand{\cR}{\mathscr{R}}
\newcommand{\cC}{\mathscr{C}}

\newcommand{\ceil}[1]{\lceil #1 \rceil}

\newcommand{\code}[1]{\ensuremath{\mathsf{#1}}}

\newcommand{\connected}[1]{\def\temp{#1}\ifx\temp\empty\sim\else\overset{#1}{\sim}\fi}

\newcommand{\Av}{\mathrm{Av}}

\tikzset{
	point/.style={circle, fill, inner sep=1.5pt},
	smallpoint/.style={point, inner sep=1.2pt},
	tinypoint/.style={point, inner sep=1pt},
	hlbox/.style={fill, {white!90!black}},
	subrect/.style={draw, fill={white!80!cyan}}, 
	msrect/.style=subrect 
}

\newtheorem{theorem}{Theorem}[section]
\newtheorem{lemma}[theorem]{Lemma}

\newtheorem{corollary}[theorem]{Corollary}

\theoremstyle{definition}

\title{Compact representations of pattern-avoiding permutations}

\author[1]{L\'aszl\'o Kozma}
\author[2]{Michal Opler\thanks{This work was co-funded by the European Union under the project Robotics and advanced industrial production (reg. no. CZ.02.01.01/00/22\_008/0004590).}}
\affil[1]{Faculty of Computer Science, TU Dresden, Germany}
\affil[2]{Czech Technical University in Prague, Czech Republic}

\date{}

\begin{document}


\maketitle

\begin{abstract}
Pattern-avoiding permutations are a central object of study in both combinatorics and theoretical computer science. In this paper we design a data structure that can store any size-$n$ permutation $\tau$ that avoids an arbitrary (and unknown) fixed pattern $\pi$ in the asymptotically optimal $\fO(n \lg{s_\pi})$ bits, where $s_\pi$ is the \emph{Stanley-Wilf limit} of $\pi$. Our data structure supports $\tau(i)$ and $\tau^{-1}(i)$ queries in $\fO(1)$ time, sidestepping the lower bound of Golynski (SODA 2009) that holds for general permutations. Comparable results were previously known only in more restricted cases, e.g., when $\tau$ is \emph{separable}, which means avoiding the patterns 2413 and 3142. 

We also extend our data structure to support more complex geometric queries on pattern-avoiding permutations (or planar point sets) such as rectangle \emph{range counting} in $\fO(\lg\lg{n})$ time. This result circumvents the lower bound of $\Omega{(\lg{n}/\lg\lg{n})}$ by P\u{a}tra\c{s}cu (STOC 2007) that holds in the general case. For \emph{bounded treewidth} permutation classes (which include the above-mentioned separable class), we further reduce the space overhead to a lower order additive term, making our data structure \emph{succinct}. This extends and improves results of Chakraborty et al.~(ISAAC 2024) that were obtained for separable permutations via different techniques. All our data structures can be constructed in linear time. 

\end{abstract}

\section{Introduction}\label{sec:intro}
A permutation $\tau = (\tau_1, \dots, \tau_n)$ \emph{contains} a permutation pattern $\pi = (\pi_1, \dots, \pi_k)$ if there are indices $i_1 < \cdots < i_k$, such that $\tau_{i_j} < \tau_{i_\ell}$ if and only if $\pi_j < \pi_\ell$, for all $j,\ell \in [k]$, i.e., the subsequence $\tau_{i_1}, \dots, \tau_{i_k}$ of $\tau$ is \emph{order-isomorphic} to $\pi_1, \dots, \pi_k$. Otherwise we say that $\tau$ \emph{avoids} $\pi$. 

Pattern-avoiding permutations arise in varied contexts and have been intensively studied for decades, both in combinatorics and in theoretical computer science (e.g., see \cite{ kitaev2011patterns,bona2022combinatorics}). For many concrete patterns $\pi$, the avoidance of $\pi$ has natural alternative interpretations. For example, in one of the earliest results in this area, Knuth~\cite[\S\,2.2.1]{knuth68} showed that $231$-avoiding permutations are exactly those that are \emph{sortable with a stack}. Monotone patterns (subsequences) are ubiquitous, e.g., in the context of the Erd\H{o}s-Szekeres theorem.  The question we ask in this paper is:
\smallskip

\quad \quad \quad \quad \emph{Can pattern-avoiding permutations be compactly represented?} 
\smallskip

One can obviously store a length-$n$ permutation $\tau$ by storing $\tau_i$ for all $i\in[n]$. This can be done using $n \ceil{\lg{n}}$ bits, essentially matching the information-theoretic lower bound. However, to turn the representation into a useful data structure, one should support, preferably in constant time, both $\tau(i)$ and $\tau^{-1}(i)$ queries (given $i$, return $\tau_i$, or given $i$, return $j$ such that $\tau_j=i$), and perhaps other queries.\footnote{We use the notation $\lg(x) = \log_2(x)$, $[k] = \{1,\dots,k\}$, and we often write permutations inline, e.g., $231$ for $(2,3,1)$. Both $\tau_i$ and $\tau(i)$ indicate the same value, but the latter notation is used when referring to a query.} In other words, we would like to optimally compress a permutation in a way that allows accessing it \emph{locally}, without fully decompressing it upon each query. 


A simple solution is to store both $\tau(i)$ and $\tau^{-1}(i)$, for all $i \in [n]$, doubling the required space. 
An improved scheme that uses $(1+\varepsilon)n\lg{n}$ bits, for any $0 < \varepsilon < 1$, was given by Munro, Raman, Raman, and Rao~\cite{MRRR}, supporting $\tau(i)$ and $\tau^{-1}(i)$, and in fact, arbitrary power-$k$ queries $\tau^k(i)$, in time $\fO(1/\varepsilon)$. Golynski~\cite{Golynski} showed this tradeoff between space and query time to be essentially optimal. 

Data structures whose space usage is optimal up to a constant factor are usually called \emph{compact}; they have been studied extensively for storing permutations, trees, graphs, vectors, and other objects (e.g., see the textbook by Navarro~\cite{book_compact} for a broad survey). 

Data structures with a stricter space requirement, having only a lower order additive overhead, are called \emph{succinct}. 
Munro et al.~\cite{MRRR} also design {succinct} schemes for representing permutations, i.e., using $n\lg{n} + o(n\lg{n})$ bits, but the query costs in this case are necessarily super-constant.  

\medskip

All the mentioned solutions are, however, too wasteful for storing pattern-avoiding permutations. As the landmark result of Marcus and Tardos~\cite{MarcusTardos} shows, the number of length-$n$ permutations that avoid a fixed pattern is \emph{single-exponential} in $n$; this has long been known as the Stanley-Wilf conjecture.\footnote{The proof of the conjecture by Marcus and Tardos~\cite{MarcusTardos} also builds on work by Füredi and Hajnal~\cite{FurediHajnal}, and Klazar~\cite{Klazar}.} This result is the main motivation for our work, as it allows, in principle, a data structure with \emph{linear} space. More precisely, denoting by $\Av_n(\pi)$ the set of length-$n$ permutations that avoid $\pi$, we have $|\Av_n(\pi)| \leq s_{\pi}^n$, for a quantity $s_\pi$ known as the \emph{Stanley-Wilf limit} of $\pi$. Note that $s_\pi$ depends only on $\pi$, and can be defined as the limit $s_\pi = \lim_{n \rightarrow \infty}{\sqrt[n]{|\Av_n(\pi)|}}$~\cite{Arratia99}.\footnote{The growth rate of $s_\pi$ as a function of $k = |\pi|$ is far from fully understood, it is known~\cite{jfox} that $s_\pi \in 2^{O(k)}$ for all $\pi$, and $s_\pi \in 2^{\Omega(k^{1/4})}$ for some $\pi$.}  

 A natural goal is then to store a $\pi$-avoiding permutation $\tau$ compactly, i.e., in the asymptotically optimal space of $\fO(\lg{|\Av_n{(\pi)}|}) = \fO(n\lg{s_{\pi}})$ bits, while supporting $\tau(i)$ and $\tau^{-1}(i)$ queries in time independent of $n$ and $\pi$. 
In addition, the data structure that stores $\tau$ should be efficient to construct (ideally, in linear time). The reader may notice, that a data structure with these properties would allow \emph{sorting} pattern-avoiding permutations in linear time. It follows from the Marcus-Tardos bound and a classical result of Fredman~\cite{Fredman} that a linear number of comparisons are sufficient for this task; however, an algorithm with actual linear running time (i.e., also including the overhead needed to find the comparisons) is far from obvious and was found only very recently~\cite{Opler_sort}. The data structure we aim for can thus be seen as a strengthening of this sorting result; to arrive at it, however, we will need a rather different set of techniques. 

\medskip


For some specific patterns $\pi$, permutations avoiding $\pi$ have rigid structure. For instance, 231-avoiding permutations are preorder-sequences of binary search trees, and this underlying tree-structure can be used rather straightforwardly to compactly represent them. 

Recently, Chakraborty, Jo, Kim, and Sadakane~\cite{ChakrabortyJ0S24} gave a succinct data structure for the slightly more general class of \emph{separable permutations}, supporting constant-time $\tau(i)$ and $\tau^{-1}(i)$ queries. Separable permutations are defined by the avoidance of the patterns $2413$ and $3142$~\cite{Bose_PPM}. These permutations also have a natural binary tree representation (they can be recursively constructed via the \emph{sum} and \emph{skew sum} operators~\cite{Bose_PPM, ChakrabortyJ0S24, STOC24})  and Chakraborty et al.~exploit this structure to build their efficient representation. 

Using related techniques, Chakraborty et al.~also design a data structure for storing \emph{Baxter-permutations}, although with polylogarithmic query times. Baxter permutations are not defined using classical pattern-avoidance, but via the related concept of \emph{vincular patterns}. Baxter permutations are in bijection with floorplans/rectangulations~\cite{Baxter1, Baxter2} and admit a representation using min-/max- Cartesian trees, which is explicitly used in~\cite{ChakrabortyJ0S24}.  

These techniques are quite specific to separable and Baxter permutations. For general pattern-avoiding permutations a straightforward hierarchical structure is not apparent, and Chakraborty et al.~leave open the question of efficiently representing other pattern-avoiding classes. Our main result addresses this question generally, for \emph{arbitrary} avoided patterns. 

\begin{theorem} \label{thm1}
Any permutation $\tau$ of size $n$ that avoids a pattern $\pi$ can be represented in $\fO(n\lg{s_\pi})$ bits, supporting $\tau(i)$ and $\tau^{-1}(i)$ queries in time $\fO(1)$.
Moreover, the representation can be constructed in $\fO(n \lg{s_\pi})$ time.
\end{theorem}


To achieve this result, we use a \emph{balanced decomposition} (partitioning or gridding) of permutations, related to merge sequences in the low \emph{twin-width} regime. A similar technique was introduced in~\cite{STOC24}, and we adapt and refine it here for the task at hand, also giving an efficient method for constructing it. 
The decomposition is fairly general, we believe that with its strong balance properties we introduce here, it may have further applications.

Our data structure also supports, in time $\fO(1)$ and with negligible overhead, a host of other queries. These include (interval) \emph{range minimum} (given $i$ and $j$, return the smallest element in $\tau_i, \dots, \tau_j$), \emph{next smaller} (given $i$, return the smallest index $j>i$, such that $\tau_j<\tau_i$), and all their symmetries.

Like other related works, our data structure uses the Word RAM model with $\Theta(\lg{n})$ word-length, space usage is however measured in bits, not words.  
We assume $\pi$ to be fixed (i.e., not depending on $n$), but we emphasize that the query times do not hide a dependence on $\pi$. In the space and construction time bound we make the optimal $\lg{s_{\pi}}$ factor explicit, and note that for the space bound the $\fO$-notation only hides a factor of two.  
Importantly, the avoided pattern $\pi$ is not known during construction and operation, and only $\tau$ is given as input. We remark that the result circumvents Golynski's lower bound~\cite{Golynski} on the time-space tradeoff attainable for general permutations.


\medskip

Reducing the constant factor in the space bound from two to one, and thus making the data structure succinct, is a challenging open question. We can achieve this for the case of \emph{bounded treewidth} permutation classes, which include separable permutations as a special case.

More precisely, to each permutation $\tau$, one can associate a graph $G_\tau$, called its \emph{incidence graph}. The vertices of $G_\tau$ are the pairs $(i,\tau_i)$ and vertex $(i,j)$ is connected to the vertices $(i+1, \tau({i+1})), (i-1, \tau({i-1})), (\tau^{-1}({j+1}),j+1), (\tau^{-1}({j-1}),j-1)$ whenever these are well-defined. The \emph{treewidth} of $G_\tau$ is then referred to as the treewidth of $\tau$. This natural complexity-measure of permutations has played an important role in pattern matching algorithms (e.g., see~\cite{AR_PPM, BKM_PPM}).

\begin{theorem}\label{thm2}
Any permutation $\tau$ of size $n$ and treewidth $t \in \fO(1)$ that avoids a pattern $\pi$ can be represented in $n\lg{s_{\pi}} + o(n)$ bits, supporting $\tau(i)$ and $\tau^{-1}(i)$ queries in time $\fO(1)$.
Moreover, the representation can be constructed in $\fO_t(n)$ time.
\end{theorem}

When $\tau$ is separable, its treewidth is known to be at most $7$~\cite{AR_PPM}, so Theorem~\ref{thm2} applies.\footnote{In fact, the more relevant complexity measure is \emph{gridwidth}, and separable permutations are exactly those of gridwidth one. It is known that treewidth and gridwidth are within a small constant factor of each other~\cite{JelinekOV18}.} Note however, that since separable permutations are defined by \emph{two} avoided patterns, the $s_\pi$ constant for either of those would not capture the best possible constant factor in the space bound. However, we can extend Theorem~\ref{thm2} to any \emph{supermultiplicative} class $\cC$ of permutations of bounded treewidth, for a space bound of $\lg{|\cC_n|} + o(n)$ where $\cC_n$ denotes the set of length-$n$ permutations in~$\cC$, thus obtaining a \emph{succinct} structure also for separable permutations. 
Here, supermultiplicative means that for any $i, j$ we have $|\cC_{i+j}| \ge |\cC_i| \cdot |\cC_j|$. 
Note: separable permutations are counted by the Schr\"{o}der numbers~\cite{West95}, so their succinct data structure uses $n \cdot \lg{(3+2\sqrt{2})} + o(n) \approx 2.54n$ bits.
As in the case of Theorem~\ref{thm1}, we can also support interval \emph{range minimum}, \emph{next smaller}, and similar queries in $\fO(1)$ time. Previously, succinctly supporting these operations was only known for the separable special case~\cite{ChakrabortyJ0S24}, where some of the operations needed $\Theta(\lg\lg n)$ time.

We note that the query times in Theorem~\ref{thm2} depend on neither $t$ nor $|\pi|$.

\paragraph{Adaptivity to pattern-avoidance.}

Our result can be seen as an algorithmic strengthening of the Stanley-Wilf, Marcus-Tardos~\cite{MarcusTardos} bound, turning it into an efficient data structure. This fits in a broader line of work studying the algorithmic consequences of pattern-avoidance, of which Knuth's stack sorting result can be seen as a first example. Fine-grained complexity bounds for pattern-avoiding inputs were obtained for several algorithmic problems. These include: searching in binary trees with rotation~\cite{FOCS15, ChalermsookPettieY, STOC24}, the $k$-server and the traveling salesman problems~\cite{STOC24}, factorization~\cite{BonnetBourneufEtAl2024},  permutation pattern matching and counting~\cite{Bose_PPM, AR_PPM, GM_PPM, BKM_PPM, GR_PPM, even2021counting}, approximate counting~\cite{ben2024approximate}, and sorting~\cite{Opler_sort}. 

In some of these results, pattern-avoidance is extended from permutations to planar point sets (under a general position assumption) in a straightforward way; this allows studying algorithmic problems on pattern-avoiding point sets.
Our data structure for storing permutations can also be extended to support geometric queries such as the natural and well studied (rectangle) \emph{range minimum}, or (rectangle) \emph{range counting}. 

In rectangle range minimum, we ask, given integers $a,b,c,d \in [n]$, for the smallest $i \in [c,d]$, such that ${\tau^{-1}}(i) \in [a,b]$.
This is more general than (interval) range minimum where only left and right boundaries are given, and where sophisticated constant-time solutions exist~\cite{rmq1,rmq2,rmq3, rmq4}.
It is also natural to consider symmetric queries, i.e., we can ask, given integers $a,b,c,d \in [n]$, for the largest $i \in [c,d]$, such that ${\tau^{-1}}(i) \in [a,b]$, or for the smallest or largest index $i \in [a,b]$, such that $\tau(i) \in [c,d]$.
In range counting, we ask, given $a,b,c,d \in [n]$, for the number of distinct indices $i \in [a,b]$, for which $\tau(i) \in [c,d]$.

\begin{theorem} \label{thm3}
Any permutation $\tau$ of size $n$ that avoids a pattern $\pi$ can be represented in $\fO(n\lg{s_\pi})$ bits, supporting rectangle range minimum and rectangle range counting queries in $\fO(\lg\lg{n})$ time.
Moreover, the representation can be constructed in $\fO(n \lg{s_\pi})$ time.
\end{theorem}

While it is convenient to state our results in terms of permutations, it is not difficult to extend them to arbitrary planar point sets, where both input points and queries are given by coordinates, e.g., as RAM words. The query times, again, are independent of the avoided pattern $\pi$.

In computational geometry, orthogonal range searching and its variants are among the best studied problems~\cite[\S\,40]{handbook_dcg}. For \emph{rectangle range counting} in the plane, classical range-tree based data structures achieve $O(n\log{n})$ space and $\fO(\log{n})$ query time.
In the Word RAM model, assuming $\Theta(\lg{n})$-bit words and polynomial-size coordinates, the query time has been improved to $\fO(\lg{n}/\lg\lg{n})$, with $\fO(n)$ words of space~\cite{Chazelle, Jaja}. P\u{a}tra\c{s}cu~\cite{Patrascu_lb} showed a matching lower bound on the query time in the \emph{cell-probe} model, even assuming $\fO(n\lg^{\fO(1)}{n})$ space. Our result circumvents this lower bound using the special properties of the input. Surprisingly, both \emph{rectangle emptiness} and \emph{rectangle range minimum} are solvable, in general point sets, with $\fO(\lg\lg{n})$ query times and $\fO(n\lg\lg{n})$, resp., $\fO(n\lg^{\varepsilon}{n})$  words of space~\cite{ChanLarsenP}; for these tasks our result only improves the space and preprocessing time bounds.

\paragraph{Sorting and compact data structures.}
Data structures for restricted classes of permutations have also been considered by Barbay and Navarro~\cite{BN1, Barbay2}. The families of inputs they study come from classical \emph{adaptive sorting}, e.g., permutations with few \emph{contiguous sorted runs} or \emph{interleaved sorted runs}, where the space- and time bounds  depend on the distribution of run lengths. 
An input that consists of $k-1$ interleaved sorted sequences clearly avoids $\pi = k,\dots,1$, our results thus subsume these cases. Moreover, permutations with few \emph{contiguous} runs form a supermultiplicative class of bounded treewidth, our results thus imply a succinct data structure in this special case.  

The connection between adaptive sorting and compact data structures observed in these works has also more broadly inspired our work. As mentioned already, efficient compact data structures yield efficient adaptive sorting algorithms; the reverse, however, is not known to hold generally, and has only been established in some special cases~\cite{Barbay2}. 
Thus, the recent linear-time sorting algorithm for pattern-avoiding permutations~\cite{Opler_sort} suggests the natural question of whether a compact data structure exists for this family of inputs, affirmatively answered by Theorem~\ref{thm1}. 

\paragraph{Bounded twin-width permutations and matrices.} 
Twin-width is a recently emerging complexity measure for permutations, graphs, as well as more general structures~\cite{GM_PPM, bonnet2021twin, bonnet2021twinperm}. Although we mostly avoid terminology related to twin-width to keep the presentation self-contained, our results and techniques closely relate to this concept.  

In particular, it is known that a permutation $\tau$ that avoids $\pi$, as well as its matrix $M_\tau$, have twin-width at most $2^{O(|\pi|)}$~\cite{GM_PPM, jfox}.  
We could therefore directly use a recent compact data structure for storing low twin-width matrices, by Pilipczuk, Soko{\l}owski, and Zych-Pawlewicz~\cite{matrix_ds}. While it represents a more general class of objects, this data structure would fall short of our goals in several aspects: its query times are $\fO(\lg\lg{n})$, supporting only a subset of the queries we consider, taking superlinear time to construct. More importantly, the $\fO$-notation in the space and query-time bounds involves an exponential dependence on twin-width; in our setting that would mean a doubly-exponential dependence on $|\pi|$. In contrast, in our data structure the query times are independent of $\pi$ and the space bound involves an optimal $\lg{s_\pi} \in \fO{(|\pi|)}$ factor. Nonetheless we adopt some terminology related to matrix-divisions from~\cite{matrix_ds}, and note some similarity in techniques, particularly in the proof of Theorem~\ref{thm3}. In turn, our results also directly extend to storing permutations of \emph{bounded twin-width} since every permutation of twin-width~$d$ avoids a certain pattern of length $(d+1)^2$~\cite{GM_PPM}.

\begin{corollary}\label{cor2}
Any permutation $\tau$ of size $n$ and twin-width $d \in \fO(1)$ can be represented in $\fO_d(n)$ bits, supporting $\tau(i)$ and $\tau^{-1}(i)$ queries in time $\fO(1)$.
Moreover, the representation can be constructed in $\fO_d(n)$ time.
\end{corollary}

A related very recent result is that $\pi$-avoiding permutations can be expressed as a composition of $\fO_{\pi}(1)$ many separables~\cite{BonnetBourneufEtAl2024}. Thus, by concatenating data structures for these separable permutations, one could represent a $\pi$-avoiding permutation, supporting some of the queries that we consider; in this way, however, the construction time, as well as space- and query time bounds would involve a \emph{doubly-exponential} dependence on $|\pi|$.
We especially stress that our data structure supports queries in $\fO(1)$ time independent of the twin-width.


\paragraph{High-level description of our techniques.} 

At a high level, our main data structure decomposes a permutation at two different levels of granularity. 
One can think of these levels as ``merging'' groups of neighboring rows or columns of the input permutation (viewed as a matrix), where the groups are of roughly equal, and carefully optimized size.
\footnote{We note that our data structure significantly differs from previous schemes for storing general permutations, that are typically based on the cycle-structure of the input.}
Due to the pattern-avoiding properties of the input, both levels of the decomposition have strong sparsity- as well as balancedness-properties (as defined in \S\,\ref{sec:prelim}, Lemma~\ref{lem:decomp}). 

We briefly describe how a $\tau(i)$ query is answered (a $\tau^{-1}(i)$ query is analogous, as essentially all parts of our data structure are also stored in a duplicate, transposed form). We first locate the vertical strips of the decomposition where $i$ falls, as well as the precise offsets within these; for this we use succinct tree and bitvector structures that allow navigating the levels of the decomposition. 

To go from column- to row-coordinates, i.e., to compute $\tau_i$ from $i$, there are multiple challenges to overcome. As a key step, we store the sub-permutations induced by each vertical strip of the decomposition; this can be done compactly due to the fact that these permutations are themselves pattern-avoiding, and thus, the overall number of distinct permutations that arise can be tightly bounded. However, collapsing a vertical strip of the matrix into a permutation loses the precise row-information, so at first we can only get partial data from this structure: the index of the cell into which the entry $(\tau_i,i)$ falls in a sparse representation of the column, and the rank of the point within the permutation induced by its cell. 

A separate data structure connects the cells at different levels of granularity and ultimately allows to identify the horizontal strips of the decomposition where $(\tau_i,i)$ falls. We convert the indices of these strips into their global vertical position by again using the succinct tree and bitvector structures. Finally, to obtain the precise vertical offset of the point \emph{within} its horizontal strip, we consult a data structure that stores the permutations within horizontal strips, built analogously to those for vertical strips. This yields the value $\tau_i$ in $\fO(1)$ time. We describe the details in \S\,\ref{secthm1}. 

\paragraph{Extensions.} 

To support the geometric queries (Theorem~\ref{thm3}), we precompute more information at each level of the decomposition, from which rectangle-statistics can be reconstructed. For the original two-level decomposition, this would require too much space. We therefore construct $\Theta(\lg\lg{n})$ levels instead, carefully controlling their granularity and the branching factor between consecutive levels. In this way, the amount of data stored at each level is kept small, while the running time is bounded as $\fO(\lg\lg{n})$. We describe the data structure in \S\,\ref{sec4}.

For permutations of bounded treewidth (Theorem~\ref{thm2}) we again use only two levels of the decomposition. However, we avoid separately storing both horizontal- and vertical strips. Instead we decompose the permutation into \emph{components} -- groups of cells in the decomposition that contain input points and that are connected by visibility. Components can reach multiple horizontal and vertical strips, and they contain \emph{all} input points in those strips. Low treewidth guarantees that they are of bounded size; the two properties together allow storing sufficient detail for each component that allows converting between row- and column-information. We refer to \S\,\ref{sec5} for details. 

\paragraph{Open questions.}
For our main result (Theorem~\ref{thm1}), we leave open the question of a factor-two improvement in the space bound, i.e., whether the scheme can be made succinct. This factor is inherent in our design, due to the need to store both column-, and row-permutations, even for answering just $\tau(i)$-queries. We also leave open whether $\tau^k(i)$ queries can be supported within similar bounds. 

For our geometric data structure (Theorem~\ref{thm3}), can the cost of range counting queries be improved to $\fO(1)$, or is there a non-trivial lower bound, or optimal trade-off result? In geometric applications, \emph{weighted} versions of range searching/counting are often considered. Incorporating weights in our solution is problematic, since we store parts of the input only up to isomorphism.  Informally, if we supported weights in full generality, then these could ``encode'' arbitrary permutations, and the classical lower bounds would apply. In a different direction, extending our results to a dynamic setting also appears challenging.

Can efficient compact representations be obtained for other interesting classes of permutations, not defined by pattern-avoidance? 
More broadly, can a compact data structure for a family $\mathcal{P}_n$ of permutations be obtained whenever sorting in $\fO(\lg{|\mathcal{P}_n|})$ time is possible, or is there a separation between the two problems for some natural $\mathcal{P}_n$? 

\section{Preliminaries}\label{sec:prelim}
\paragraph{Permutation pattern-avoidance.}

Let $\tau = (\tau_1, \dots, \tau_n)$ be a permutation. We call $\tau$ \emph{non-trivial} if $n>1$. We can bijectively map $\tau$
to an $n \times n$ (permutation) matrix $M_\tau$ where the $\tau_i$-th entry of the $i$-th column is $1$ for all $i \in [n]$ and all other entries are $0$. We refer to matrix entries by row first and column second, ordering top to bottom and left to right. We also frequently refer to a $1$-entry $(\tau_i,i)$ as a \emph{point}. 
The statements ``matrix $M_\tau$ contains/avoids matrix $M_\pi$'' 
are interpreted naturally to mean ``$\tau$ contains/avoids $\pi$''. 

A general $0$-$1$ matrix $M$~\emph{contains}~$M_\pi$ or simply \emph{$M$~contains~$\pi$}, if $M_\pi$ can be obtained from $M$ by deleting rows and columns and changing $1$-entries to $0$.  Otherwise we say that $M$~\emph{avoids}~$M_\pi$ or simply \emph{$M$~avoids~$\pi$}.

For a permutation $\pi$, we denote by $\ex_{\pi}(n)$, the maximum number of ones in an $n \times n$ $0$-$1$ matrix that avoids $\pi$. Marcus and Tardos~\cite{MarcusTardos} showed that for every fixed $\pi$, the quantity $\ex_{\pi}(n)$ is linear in $n$. More precisely, there exists a quantity $c_\pi$ so that:

\begin{lemma}\label{lem:mt}
Every $n \times n$ $0$-$1$ matrix $M$ with at least $c_\pi \cdot n$ one-entries contains $\pi$. 
\end{lemma}
The optimal value $c_{\pi}$ is known as the \emph{F\"{u}redi-Hajnal limit} of $\pi$, after the earlier conjecture~\cite{FurediHajnal}, and can be defined as $c_{\pi} = \lim_{n \rightarrow \infty}{\ex_{\pi}(n)/n}$.  It is known that the F\"{u}redi-Hajnal limit and the Stanley-Wilf limit (defined in \S\,\ref{sec:intro}) are polynomially related; concretely, Cibulka~\cite{Cibulka2009} showed $s_{\pi} = \Omega(c_{\pi}^{2/9}) \cap \fO(c_{\pi}^2)$. 

\begin{figure}
\centering\includegraphics[scale=0.35]{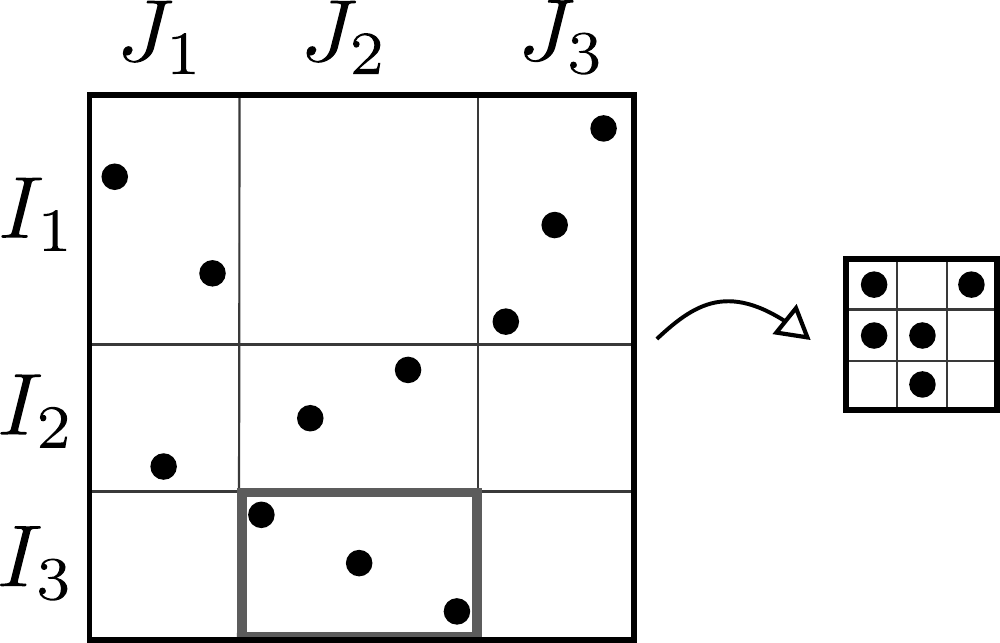}
		\caption{Permutation matrix $M$ of size $11 \times 11$ (left) with division $(\cR,\cC)$ and $M(\cR,\cC)$ (right), with $\cR = (I_1,I_2,I_3)$, $\cC = (J_1,J_2,J_3)$, and cell $M[I_3,J_2]$ framed. Here, $I_1 = [1,5], J_1 = [1,3], I_2 = [6,8], J_2=[4,8], I_3 = J_3 = [9,11]$. Dots correspond to $1$-entries, empty spaces to $0$-entries.}\label{fig1}
        \end{figure}

\paragraph{Divisions and coarsening.} Let $M$ be an $n \times n$ $0$-$1$ matrix. 
Let $\cR = (I_1, \dots, I_k)$ be a partition of $[n]$, i.e., $I_1 \cup \cdots \cup I_k = [n]$ and $I_1, \dots, I_k$ are pairwise disjoint \emph{contiguous intervals}. 
Similarly, let $\cC = (J_1, \dots, J_{k'})$ be a partition of $[n]$. 
Then, we refer to $(\cR, \cC)$ as a \emph{division} of $M$; intuitively, the intervals in $\cR, \cC$ capture neighboring sets of rows and columns in $M$ that are \emph{merged} together. 
The resulting matrix $M' = M(\cR,\cC)$ is defined as follows. Assuming $\cR, \cC$ as above, $M'$ has $k$ rows and $k'$ columns, and entry $M'(i,j)$ is $1$ if at least one entry $M(x,y)$ with $x \in I_i$ and $y \in J_j$ equals $1$, and $M'(i,j)$ is $0$ otherwise. When clear from the context, by the $i$-th row, resp., $j$-th column of the division $(\cR,\cC)$ we refer to the submatrix of the original matrix $M$ defined by the rows $I_i$, resp., by the columns $J_j$. Accordingly, the \emph{height} of the $i$-th row is $|I_i|$, i.e., the number of rows of the original matrix that are merged into row $i$ of $M'$. Similarly, the \emph{width} of the $j$-th column is $|J_j|$. By $M[I_i,J_j]$ we refer to the submatrix of $M$ at the intersection of rows indexed by $I_i$ and columns indexed by $J_j$. We also call this submatrix the \emph{cell} $(i,j)$ of the division $(\cR,\cC)$. A cell is non-zero, if it has at least one non-zero entry. We illustrate some of these concepts in Figure~\ref{fig1}. 

We say that a division $(\cR', \cC')$ is a \emph{coarsening} of $(\cR, \cC)$ if $\cR'$ can be obtained from $\cR$ by successively replacing neighboring intervals by their union, and if $\cC'$ can be similarly obtained from $\cC$. In this case, we also say that $(\cR, \cC)$ is a \emph{refinement} of $(\cR', \cC')$. Notice that divisions form a partial order by the coarsening relation, where the division with a single $[n]$ interval for both rows and columns is the coarsest, and the division with singleton sets for both rows and columns (i.e., capturing the original matrix) is the finest (i.e., least coarse).

\paragraph{Data structuring tools.}
We use as building blocks the following compact data structures. Note that all these structures are static: once built, they are not modified. While we omit discussing the details of their construction, this takes linear time for all components and is subsumed by the $\fO(n \lg{s_\pi})$ time necessary for sorting the input $\tau$.

\begin{itemize}
\item \emph{sparse bitvector}: stores values $B[1], \dots, B[n] \in \{0,1\}$, supporting in $\fO(1)$ time the operations:
\begin{itemize}
\item $\sread(B,i)$: returns $B[i]$,
\item $\srank(B,i)$: number of $1$-bits in $B[1, \dots, i]$,  
\item $\sselect(B,i)$: smallest $j$ such that the number of $1$-bits in $B[1, \dots, j]$ is $i$, i.e., the index of the $i$-th $1$.

\end{itemize}
The data structure uses $n \mathcal{H} +o(n)$ bits, where $\mathcal{H}$ is the \emph{empirical entropy} of the bitvector, i.e., $\mathcal{H} = p\lg{\frac{1}p} + (1-p)\lg{\frac{1}{1-p}}$, where $p$ is the fraction of $1$-bits in $B[1,\dots,n]$; notice that $0 \leq \mathcal{H} \leq 1$. 
For possible implementations with these parameters, we point to~\cite[\S\,4]{book_compact} and references therein. 

\item \emph{tree}:
rooted, ordered tree with~$n$ nodes, supporting the following operations in~$\fO(1)$ time:
\begin{itemize}
\item $\code{parent}(v)$: parent of node~$v$,
\item $\code{child}(v,i)$: $i$-th child of node~$v$,
\item $\code{childRank}(v)$: number of siblings to the left of node~$v$,
\item $\code{leafSelect}(i)$: $i$-th leaf from the left,
\item $\code{leafRank}(v)$: number of leaves to the left of node~$v$, excluding the leaves in subtree of $v$. 
\end{itemize}
Standard implementations using $2n + o(n)$ bits can be found in \cite[\S\,8]{book_compact} and references therein. 

\item \emph{function}: a mapping $A \rightarrow B$, with $A = [|A|]$ and $B=[|B|]$, where $|B|$ is assumed to fit in a machine word. The space requirement of a standard implementation is $|A| \cdot \lceil \lg{|B|} \rceil$ bits. (A stronger bound of $\lceil |A| \cdot \lg |B| \rceil + \fO(1)$ bits is achievable, but not needed in our work.)

\item \emph{range minimum} data structure: processes an array $A[1, \dots, n]$, so that, upon queries $\code{rm}(a,b)$ for $a,b \in [n]$, it returns an index $i \in [a,b]$ for which $A[i] = \min{A[a,\dots, b]}$. (For simplicity we assume array entries to be distinct.) Standard solutions are based on an equivalence between range-minimum queries over $A$ and \emph{lowest common ancestor} queries in the Cartesian tree built from $A$, e.g., see~\cite{rmq1,rmq2,rmq3,rmq4}.
It is possible to answer $\code{rm}(a,b)$ queries in $\fO(1)$ time, with a data structure using $2n + o(n)$ bits of space; crucially, the data structure can be implemented so that the array $A$ does not need to be read during queries. 



\end{itemize}

\paragraph{Main structural lemma.}
An important ingredient of our data structure is the following decomposition of a permutation (matrix). As mentioned, this decomposition is related to merge sequences of low twin-width permutations (e.g., \cite{GM_PPM}), significantly adapted for our purposes.

\begin{lemma}[Balanced Decomposition]\label{lem:decomp}
	Let $\pi$ be a non-trivial permutation with Füredi-Hajnal limit $c_\pi$ and let $M$ be a $\pi$-avoiding $n\times n$ permutation matrix.
	
	For integer parameters $1 < m_1 < m_2 < \dots<  m_\ell < n$, there exist divisions $(\cR_1, \cC_1), \dots, (\cR_\ell, \cC_\ell)$ of~$M$ such that:
	\begin{enumerate}[(a)]
		\item The division~$(\cR_i, \cC_i)$ is a coarsening of $(\cR_{i+1}, \cC_{i+1})$ for each $i \in [\ell-1]$.
		\item 
        $M(\cR_i, \cC_i)$ 
        has exactly $m_i$ rows and columns.\label[ipart]{item:bal-gridding:num-rows}
		\item Each row in~$\cR_i$ and each column in $\cC_i$ has height, resp., width at most $40 \cdot n/m_i$.\label[ipart]{item:bal-gridding:width-rows}
		\item Each row and each column in $M(\cR_i,\cC_i)$ has at most 
        $10c_\pi$ non-zero 
        cells.
        \label[ipart]{item:bal-gridding:points-per-row}
		\item Each row in~$\cR_i$ and each column in $\cC_i$ is obtained from merging at most $40 \cdot m_{i+1}/m_i$ rows of $\cR_{i+1}$, resp., columns of~$\cC_{i+1}$, 
        for $i \in [\ell-1]$. 
	\end{enumerate}
	Moreover, the divisions $(\cR_1, \cC_1), \dots, (\cR_\ell, \cC_\ell)$ can be computed in time $\fO((\lg c_\pi + 1) \cdot n + c_\pi \cdot\sum_{i=1}^{\ell}m_i)$ even if $\pi$ and its Füredi-Hajnal limit~$c_\pi$ are a priori unknown.
\end{lemma}

We give a self-contained proof of Lemma~\ref{lem:decomp} in \S\,\ref{sec6}. A weaker statement of a similar flavor was obtained in~\cite{STOC24}. In particular, Lemma~\ref{lem:decomp} differs in enforcing an exact number of columns and rows in part (b), an additional balancedness condition between the layers in part (e) and very efficient implementation.

\section{Compact data structure for pattern-avoiding permutations}\label{secthm1}

In this section, we describe the main data structure referred to in Theorem~\ref{thm1}. Let $\tau$ be an input permutation of length $n$, and let $M = M_\tau$ be its corresponding permutation matrix. 
 To simplify presentation, we omit floor and ceiling operators, and we assume that $n$ is larger than some unspecified constant.

The data structure is based on two non-trivial levels of the balanced decomposition (Lemma~\ref{lem:decomp}). More precisely, we set $\ell = 2$ and find a decomposition with the parameters $1 < m_1 < m_2 < n$ where $m_1 =  n/\lg^2{n} $ and $m_2 =  n/\sqrt{\lg{n}} $. We will sometimes refer to the resulting divisions $(\cR_1,\cC_1)$ and $(\cR_2,\cC_2)$ as the \emph{coarse}-, resp., \emph{fine} division, and we refer to their cells as $1$-cells, resp., $2$-cells.

\begin{figure}
\centering\includegraphics[scale=0.57]{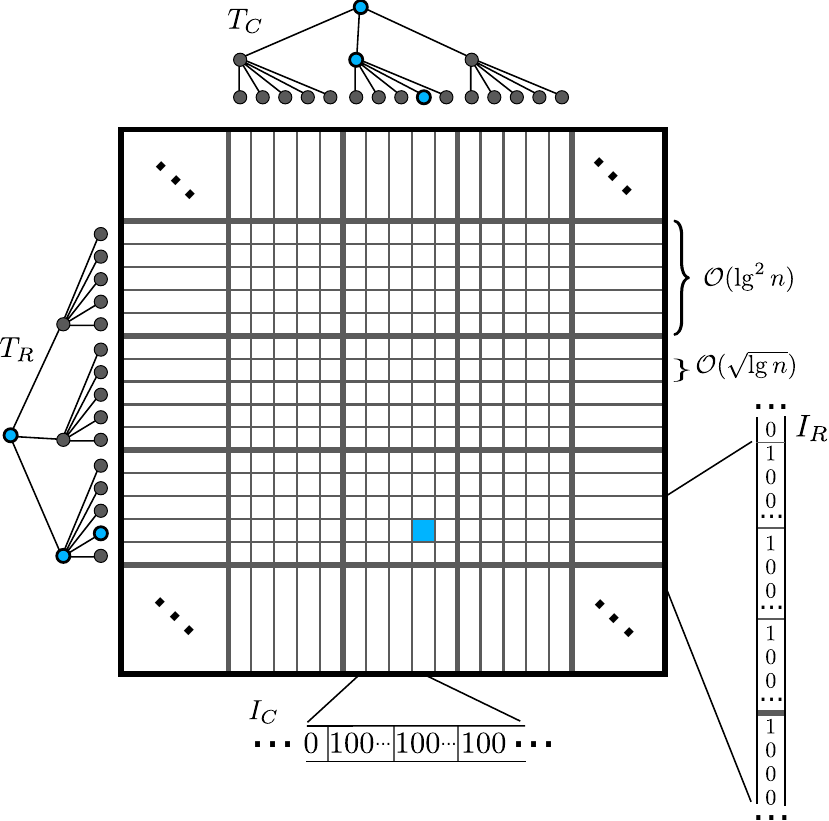}
		\caption{Decomposition of an $n \times n$ permutation matrix $M$ with a (coarse) division $(\cR_1,\cC_1)$ that is a coarsening of a (fine) division $(\cR_2,\cC_2)$. Column and row trees $T_C, T_R$ and column and row indices $I_C, I_R$ illustrated. A $2$-cell and corresponding nodes in $T_R$ and $T_C$ highlighted.}\label{fig_comp1}
        \end{figure}

\begin{figure}
\centering\includegraphics[scale=0.23]{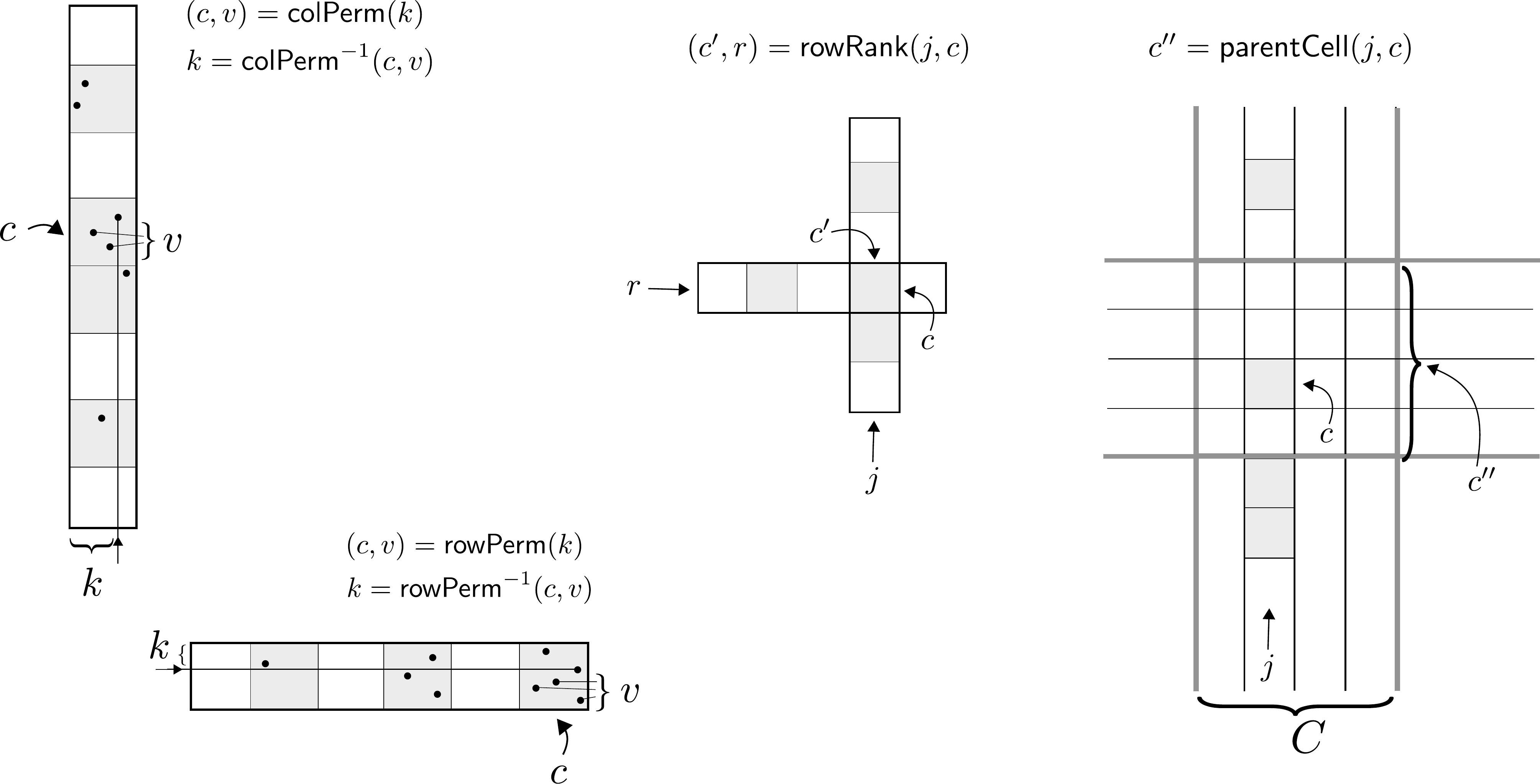}
		\caption{\emph{Left}: column and row permutation mapping and their inverse. Value $k$ is the absolute offset of the query within the column or row; $c$ refers to the $c$-th non-zero cell within column or row (non-zero cells are shaded gray); $v$ is the vertical rank of the query within its cell, i.e., $v$ entries in cell are below query point.\\
        \emph{Right}: $\code{rowRank}$ method: the $c$-th non-zero cell of the $j$-th column is the $c'$-th non-zero cell of the $r$-th row; 
         $\code{parentCell}$ method invoked for a $\cC_1$-column $C$: the $c$-th non-zero $2$-cell in the $j$-th $\cC_2$-column in $C$ is in the $c''$-th non-zero cell of $C$.}\label{fig_comp2}
        \end{figure}

We now list the components and their space requirements. See Figures~\ref{fig_comp1} and \ref{fig_comp2} for illustration.

\paragraph{A. Column and row trees $T_{C}, T_R$.} The \emph{column tree} $T_C$ and the \emph{row tree} $T_R$ capture the merges of the rows, resp.\ columns that yield the divisions of the balanced decomposition. 

More precisely, the root of $T_C$ corresponds to $[n]$, i.e., the set of all columns, the $m_1$ children of the root (level-1) correspond to the columns in the coarse division $(\cR_1, \cC_1)$, the $m_2$ nodes on level-2 of the tree correspond to the columns in the fine division $(\cR_2, \cC_2)$, where if node $x$ is a parent of node $y$, then the column of $x$ contains the column of $y$, and the ordering of siblings is the same as the ordering of columns in the matrix. Note that the $n$ columns of $M$ are not explicitly stored in the tree. 
The construction of the row tree $T_R$ is entirely symmetric. 

Both $T_C$ and $T_R$ are represented using the succinct tree implementation mentioned in \S\,\ref{sec:prelim}, allowing rich navigation queries.  
The number of nodes in both trees is $1+m_1+m_2$, the space requirement is thus $\fO(n/\sqrt{\lg{n}}) \subseteq o(n)$.

\paragraph{B. Column and row indices $I_C, I_R$.}

The \emph{column index} $I_C$ is a bitvector over the $n$ columns of the matrix $M$, with $1$-entries for the start of each column in the fine division, i.e., $I_C[i]=1$ if a column of $\cC_2$ begins at the $i$-th column of $M$, and $I_C[i]=0$ otherwise. Given an index $i \in [n]$ of a column in $M$, we can find the index of the $\cC_2$ column containing it using a $\srank$ query over~$I_C$. Conversely, given a column index in $\cC_2$, we can find its start index in $M$ with a $\sselect$ query over~$I_C$. Analogously, we store a bitvector $I_R$ over rows of the original matrix $M$, marking the start of each row in $\cR_2$.

Both bitvectors have size $n$ with $m_2 = n/\sqrt{\lg{n}}$ non-zero entries, their space requirement using the implementation mentioned in \S\,\ref{sec:prelim} is thus $o(n)$.

\paragraph{C. Column permutations $\code{colPerm}$.} We store, for each $\cC_2$-column~$C$ of $M$ the \emph{permutation} of the $|C|$ points of $M$ that fall into $C$. 
Although not sufficient to get the exact row position for the entry in column $i$ (which would immediately solve the $\tau(i)$ query), it allows to identify the $\cR_2$-row index of the queried point, and relative rank information within the $2$-cell where the query falls.

More precisely, the method $\code{colPerm}$ invoked on a $\cC_2$-column maps a column offset $k$ to an index~$c$ (the non-zero cell index where the query falls), and a value $v$, the number of $1$-entries that fall into the same cell $c$, below the queried point $(\tau_i,i)$. See Figure~\ref{fig_comp2}.

The cell index $c$ refers to the $c$-th non-zero cell in the column, its domain is thus $[10c_\pi]$, by Lemma~\ref{lem:decomp}(d). We also store the inverse mapping denoted $\code{colPerm}^{-1}$ that maps $(c,v)$ to $k$.

Recall that there are $m_2=n/\sqrt{\lg{n}}$ columns in $\cC_2$ and the column width is at most $w = 40\sqrt{\lg{n}}$ by Lemma~\ref{lem:decomp}(c).
Storing $\code{colPerm}$ explicitly for each column would be too costly, we therefore implement a scheme that (i) stores the \emph{permutation} of $1$-entries that fall into the column, together with the partitioning of entries into cells, and (ii) uses global precomputed tables for each permutation up to isomorphism, for space efficiency.

The number of possible permutations of points induced by $1$-entries in a column is $\sum_{k=1}^w{s_{\pi}^k} \in \fO(s_{\pi}^w)$. (We use the fact that the permutation induced by the column is also $\pi$-avoiding.) The permutation is augmented with data about how many of the (up to) $w$ entries fall into each of the (up to) $10c_{\pi}$ non-zero cells of the column. We refer to the permutation together with its augmentation as a \emph{gridded permutation}. Notice that the number of possible distinct gridded permutations is $\fO(s_{\pi}^w \cdot w^{10c_{\pi}})$.

Let $w_C = |C|$ denote the width of column $C$ ($w_C \leq w$). 
To implement $\code{colPerm}$ for $C$, we store the \emph{index} to the gridded permutation of length~$w_C$ in the column, this requires $\lceil \lg{s_{\pi}^{w_C}} \rceil + \lceil \lg{w^{10c_{\pi}}} \rceil \leq w_C \lg{s_\pi} + 10c_{\pi} \lg{w} + 2 \in w_C \lg{s_\pi} + \fO_\pi(\lg\lg n)$ bits (the first term is for the permutation, and the second for the gridding augmentation). In total, over all columns, this adds up to $n \lg{s_\pi} + o(n)$ bits, which will be accounted for in part D.



While for ease of notation we will simply write, e.g., $(c,v) \gets \code{colPerm}(k)$ or $k \gets \code{colPerm}^{-1}(c,v)$, this should be interpreted as obtaining a pointer to a permutation data structure stored in a precomputed table where each distinct gridded permutation is stored at most once. 

Although the number of possible gridded permutations to store is $\fO(s_\pi^w \cdot w^{10c_{\pi}})$, we still cannot easily compute them upfront, not knowing $\pi$. However, we can examine each column $C$ of $\cC_2$ and only store the gridded permutations that actually appear in a column. First we can bucket permutations by size (i.e., the column width $w_C$), then by permutation order class (we can identify this by using the efficient pattern-avoiding-sorting~\cite{Opler_sort}), and finally by gridding. Storing gridded permutations of the same size in separate tables ensures that the index lengths correspond to the column widths and the above space bound holds.
For all gridded permutations that appear, we build a simple data structure that explicitly stores the answer to every possible query on them, allowing $\fO(1)$-time operations. 

Recall that we need to map column offset $k$ to cell index $c$ and vertical rank $v$ of the entry within its cell.
This mapping has domain $[w]$ and range $[10c_{\pi}] \times [w]$. Storing it explicitly thus requires $\fO(w \lg (c_{\pi}w))$ bits, and $\fO(c_{\pi}w \lg {w})$ for its inverse mapping. 
As we have at most $\fO(s_{\pi}^w w^{10c_{\pi}})$ such structures, their total space requirement is  $\fO(s_{\pi}^w w^{10c_{\pi}} \cdot (w \lg (c_{\pi}w) + c_{\pi}w \lg{w})) \subseteq o(n)$.

Additionally, observe that for any permutation class $\cC$ that is a subclass of $\pi$-avoiding permutations, the scheme described encodes a column of width $w'$ in $\lg |\cC_{w'}| + \fO_\pi(\lg\lg n)$ bits where $\cC_{w'}$ is the set of all permutations in~$\cC$ of length exactly~$w'$.
This is a corollary of our approach that is oblivious to and independent of~$\pi$ except for the size of gridding given by the balanced decomposition. Over all columns, this adds up to at most $\lg{|\cC_n|} + o(n)$ bits, assuming that $\cC$ is supermultiplicative. 


\paragraph{D. Recursive column structure $G_C$.} 

This component is the crucial part of the data structure that allows converting from column to row data. It is a tree structure over the columns similar to $T_C$, but storing more detailed column-information from the input permutation matrix $M$. 

Similarly to $T_C$, the root of $G_C$ corresponds to $[n]$, i.e., the set of all columns, and the $m_1$ children of the root (level-1) correspond to the columns in the coarse division $(\cR_1, \cC_1)$, and the $m_2$ nodes on level-2 correspond to columns in the fine division $(\cR_2, \cC_2)$. 
As we need to store other information at the nodes, we do not provide the flexible navigation methods that $T_C$ offers. Instead, we only navigate the tree top-down, with the query $\code{subColumn}(j)$ that gives a pointer to the $j$-th subcolumn of the current node. More precisely, for the root, $\code{subColumn}(j)$ points to the $j$-th column in $\cC_1$, and for a level-1 node corresponding to a $\cC_1$-column, $\code{subColumn}(j)$ points to the column in $\cC_2$ that is the $j$-th within the current $\cC_1$-column.  
Level-2 nodes (leaves) do not allow $\code{subColumn}$ queries.

The key components of the interface are the methods $\code{rowRank}(j,c)$ and $\code{parentCell}(j,c)$, supported at both the root and level-1 nodes of the tree structure, and the method $\code{colPerm}$, supported at the leaf (level-2) nodes; see Figure~\ref{fig_comp2}. 

The first of these, $\code{rowRank}(j,c)$ refers to the $c$-th non-zero cell of the $j$-th subcolumn of the column associated to the current node. 
Recall that both $\cC_1$- and $\cC_2$-columns consist of cells, of which at most $10c_{\pi}$ are non-zero, and we identify the $c$-th such non-zero cell. The call $\code{rowRank}(j,c)$, returns a pair $(c',r)$ that give row-information about the identified cell. 
More precisely, when called at the root, we find out that the identified $1$-cell is the $c'$-th non-zero cell in the $r$-th $\cR_1$-row. When called at a level-$1$ node, we find out that the identified $2$-cell is the $c'$-th non-zero cell in its $\cR_2$-row, and that this row is the $r$-th among the $\cR_2$-rows within its parent $\cR_1$-row.     


The second method, $\code{parentCell}(j,c)$ similarly identifies the $c$-th non-zero cell of the $j$-th subcolumn of the column associated to the current node, and returns $c''$, indicating that the coarser cell containing the current cell is the $c''$-th non-zero cell within its column. 
Notice, that this method is only useful when called at a level-$1$ node, where we learn about the $1$-cell containing the current $2$-cell. If called at the root, the coarser cell containing the $1$-cell is necessarily the entire matrix.




Finally, for level-2 nodes (i.e., columns of the fine division $\cC_2$) we support the operation $\code{colPerm}$ that returns a pointer to a column permutation structure (as described in part~C).

\medskip

Now we describe a space-efficient implementation of $G_C$ in a recursive manner, starting from the level-2 nodes (leaves). Each level-2 node corresponds to a $\cC_2$-column~$J$ and stores its width~$|J|$ in $\fO(\lg\lg n)$ bits followed by the index into the precomputed table described in part~C.
The methods $\code{colPerm}$ and $\code{colPerm}^{-1}$ simply read off the precomputed values from the corresponding table entry in constant time.
The total space is thus $\lg s_\pi \cdot |J| + \fO_\pi(\lg\lg n)$.


The data structure of level-1 nodes consists of three parts: (i) precomputed $\code{rowRank}$ and $\code{parentCell}$ queries, (ii) data needed to implement the $\code{subColumn}$ method, and (iii) a concatenation of the data structures of its $\cC_2$ subcolumns as bitstrings.
Notice that each $\cC_1$-column contains at most $q = 40\lg^2{n}/\sqrt{\lg{n}} \in \fO(\lg^{3/2}{n})$ columns of $\cC_2$, and each $\cR_1$-row contains at most $q$ rows of $\cR_2$, by Lemma~\ref{lem:decomp}(c). 

Let us start by bounding the number of bits taken by the data structures of level-2 subcolumns (part (iii)).
Since each level-2 column $J$ takes $\lg s_\pi \cdot |J| + \fO_\pi(\lg\lg n)$ bits, this adds to $\lg s_\pi \cdot w + \fO_\pi(\lg^{3/2} n \cdot \lg\lg n) \subseteq \lg s_\pi \cdot w + \fO_\pi(\lg^{7/4} n)$ where $w$ is the width of the represented column.
For (i), to store all $\code{rowRank}(j,c)$ queries explicitly, we need to store a mapping $[q] \times [10c_\pi] \rightarrow [q] \times [10c_\pi]$ which requires $\fO({\lg^{3/2}{n}} \cdot 10c_\pi \cdot \lg{(\lg{n} \cdot 10c_{\pi}})) \subseteq \fO_\pi(\lg^{7/4} n)$ bits. Similarly, storing all $\code{parentCell}(j,c)$ queries explicitly takes $\fO_\pi(\lg^{7/4} n)$ bits. 

In order to support the $\code{subColumn}$ method, part (ii) of the node simply stores offsets to the beginning of each $\cC_2$-subcolumn (part (iii)) in an array.
Crucially, these offsets can be stored in a small number of bits since the concatenation of the data structures for level-2 subcolumns takes at most $\fO(\lg s_\pi \cdot \lg^2 n)$ bits.
Thus, each offset is stored in $\fO_\pi(\lg \lg n)$ bits, for a total of $\fO_\pi(\lg^{3/2} n \cdot \lg\lg n) \subseteq \fO_\pi(\lg^{7/4} n)$.
Altogether, this representation of a level-1 column fits into $\lg s_\pi \cdot w + \fO_\pi(\lg^{7/4} n)$ bits.

The final data structure for the root node is essentially the same.
It again consists of precomputed $\code{rowRank}$ queries, offsets needed for access to subcolumns and a concatenation of the data structures of all $\cC_1$ columns. (Recall that $\code{parentCell}$ queries at the root are not needed.)
To support $\code{rowRank}(j,c)$ queries at the root, we need to store a mapping $[m_1] \times [10c_\pi] \rightarrow [m_1] \times [10c_\pi]$. This requires $\fO_\pi(n/\lg^2{n} \cdot \lg n ) \subseteq o(n)$ bits.
The total space taken by the representations of level-1 columns is bounded by $\lg s_\pi \cdot n + \fO_\pi(m_1 \cdot \lg^{7/4} n) \subseteq \lg s_\pi \cdot n + o(n)$.
Finally, each offset needed for $\code{subColumn}$ queries takes $\fO(\lg n)$ bits for a total of $\fO_\pi(n/\lg^2{n} \cdot \lg{n}) \subseteq o(n)$.
Therefore, the complete data structure representing gridded columns recursively takes $\lg s_\pi \cdot n + o(n)$ bits. We note that this component is the bottleneck of the entire data structure.

%
%
%
%

\paragraph{E. Recursive row structure $G_R$ and row permutations $\code{rowPerm}$.}

This component is analogous to the column structure described in parts C and D. We similarly implement a method $\code{subRow}(j)$ that gives a pointer to the $j$-th subrow of the root or a level-1 row node, $\code{colRank}(j,c)$, that accesses the $c$-th non-zero cell of the $j$-th subrow, and returns $(c',s)$, where $c'$ is the identifier of the current cell in its column (among non-zero cells), and $s$ is the relative rank of the column containing the cell. Similarly to part $D$, the method $\code{parentCell}(j,c)$ returns a value $c''$, indicating that the $c$-th non-zero cell of the $j$-th subrow is contained in the $c''$-th non-zero cell of the current row.  

The row permutations $\code{rowPerm}$ are stored analogously to the column permutations. 
We write $(c,v) \gets \code{rowPerm}(\ell)$ or $\ell \gets \code{rowPerm}^{-1}(c,v)$ to indicate the mapping between $c$, the non-zero index of a cell within its row, a row offset $\ell$, and a value $v$, the number of $1$-entries that fall into the same cell $c$, below the queried entry. Again, we write $\code{rowPerm}$ and $\code{rowPerm}^{-1}$ as methods only for notational simplicity; these should be thought of as pointers to a global precomputed table from which the respective queries can be read out. To avoid duplication, we omit a more detailed description of this structure. 


Similarly to C and D, the total space requirement is $n\lg{s_{\pi}} + o(n)$. The fact that we store both column and row permutations is what leads to the factor two in the leading term of the space bound.

\paragraph{Implementation of $\tau(i)$ and $\tau^{-1}(i)$ queries.}

We now describe the implementation of the main operations. The rank and unrank ($\tau(i)$ and $\tau^{-1}(i)$) queries are symmetric. We give the pseudocode for both in Figure~\ref{alg:rank} and a detailed textual description for the first. 

\begin{figure}
\small
	\textbf{Input:} $i$\\
    \textbf{Output:} $\tau(i)$
	\begin{algorithmic}
		\State $s \gets I_C.\srank(i)$ \Comment{Find column rank}
        \State $s_i \gets I_C.\sselect(s)$ \Comment{Column start index}
		
        \State $s_3 \gets i-s_i$ \Comment{Offset within column}
        \State $\mathit{col} \gets T_C.\code{leafSelect}(s)$ \Comment{Find relative column ranks}
        \State $s_2 \gets T_C.\code{childRank}(\mathit{col})$ 
		\State $s_1 \gets T_C.\code{childRank}(T_C.\code{parent}(\mathit{col}))$
        \State $\mathit{root} \gets G_C.\code{root}$ \Comment{Navigate column structure}
\State $\mathit{col}_1 \gets \mathit{root}.\code{subColumn}(s_1)$ 
\State $\mathit{col}_2 \gets \mathit{col}_1.\code{subColumn}(s_2)$

		\State $(c_2,v) \gets \mathit{col}_2.\code{colPerm}(s_3)$ \Comment{Cell id and rank-within-cell read from column permutation}
	      \State $(c'_2, r_2) \gets \mathit{col}_1.\code{rowRank}(s_2,c_2)$ \Comment{Cell id and relative row rank}
          \State $c_1 \gets \mathit{col}_1.\code{parentCell}(s_2,c_2)$ \Comment{Parent cell id}
        \State $(c'_1,r_1) \gets \mathit{root}.\code{rowRank}(s_1,c_1)$ \Comment{Find coarse row}
         \State $\mathit{root} \gets G_R.\code{root}$ 
\Comment{Navigate row structure}
        \State $\mathit{row}_1 \gets \mathit{root}.\code{subRow}(r_1)$ 
\State $\mathit{row}_2 \gets \mathit{row}_1.\code{subRow}(r_2)$ 
\State $r_3 \gets \mathit{row}_2.\code{rowPerm}^{-1}(c_2',v)$ \Comment{Column offset read from row permutation}

\State $row \gets T_R.\code{child}(T_R.\code{child}(T_R.\code{root},r_1),r_2)$
\Comment{Find fine row}
\State $r \gets T_R.\code{leafRank}(\mathit{row})$
        \State $r_i \gets I_R.\srank(r)$ \Comment{Row start index}
        
        \textbf{return} $r_i +r_3$
        
	\end{algorithmic}


\bigskip
    \hrule
\bigskip

	\textbf{Input:} $i$\\
    \textbf{Output:} $\tau^{-1}(i)$
	
    \begin{algorithmic}
		\State $r \gets I_R.\srank(i)$ \Comment{Find row rank}
        \State $r_i \gets I_R.\sselect(r)$
		\Comment{Row start index}
        \State $r_3 \gets i-r_i$ \Comment{Offset within row}
        \State $\mathit{row} \gets T_R.\code{leafSelect}(r)$ \Comment{Find relative row ranks}
        \State $r_2 \gets T_R.\code{childRank}(\mathit{row})$ 
		\State $r_1 \gets T_R.\code{childRank}(T_R.\code{parent}(\mathit{row}))$
         \State $\mathit{root} \gets G_R.\code{root}$ \Comment{Navigate row structure}
\State $\mathit{row}_1 \gets \mathit{root}.\code{subRow}(r_1)$ 
\State $\mathit{row}_2 \gets \mathit{row}_1.\code{subRow}(r_2)$
   
		\State $(c_2',v) \gets \mathit{row}_2.\code{rowPerm}(r_3)$ \Comment{Cell id and rank-within-cell read from row permutation}
             \State $(c_2, s_2) \gets \mathit{row}_1.\code{colRank}(r_2,c_2')$ \Comment{Cell id and relative column rank}
             \State $c_1 \gets \mathit{row}_1.\code{parentCell}(r_2,c_2')$ \Comment{Parent cell id}
        \State $s_1 \gets \mathit{root}.\code{colRank}(r_1,c_1')$ \Comment{Find coarse column}
         \State $\mathit{root} \gets G_C.\code{root}$ 
\Comment{Navigate column structure}
        \State $\mathit{col}_1 \gets \mathit{root}.\code{subColumn}(s_1)$ 
\State $\mathit{col}_2 \gets \mathit{col}_1.\code{subColumn}(r_2)$
        
    \State $s_3 \gets \mathit{col}_2.\code{colPerm}^{-1}(c_2,v)$ \Comment{Row offset read from column permutation}

\State $col \gets T_C.\code{child}(T_C.\code{child}(T_C.\code{root},s_1),s_2)$
\Comment{Find fine column}
\State $s \gets T_C.\code{leafRank}(\mathit{col})$
        \State $s_i \gets I_C.\srank(s)$ \Comment{Column start index}
        
        \textbf{return} $s_i +s_3$

	\end{algorithmic}

	\caption{Implementation of the rank and unrank queries.}\label{alg:rank}
\end{figure}

Given $i$, i.e., a column index of the (unknown) permutation matrix $M_\tau$, we find using $\srank$ and $\sselect$ over $I_C$ the starting index $s_i \in [n]$ of the column in $\cC_2$ where $i$ falls, the index $s$ of this column within $\cC_2$ as well as the relative position (offset) $s_3$ within this column (clearly, $s_i+s_3 = i$). Note that if the columns of the division were of equal width, then we could compute these values by simple arithmetic, this, however is not guaranteed. The index $3$ is meant to evoke the third (trivial) level of the hierarchy. 

The column in $\cC_2$ that contains $i$ corresponds to the $s$-th leaf of the tree $T_C$. Using the tree interface, we find the rank $s_2$ of this leaf among its siblings, and the rank $s_1$ of its parent among its siblings. (In other words, the (fine) column containing $i$ is the $s$-th in $\cC_2$, and the $s_2$-th within the coarser column of $\cC_1$ that contains it. In turn, the coarser column is the $s_1$-th in $\cC_1$.)

We next navigate in the column structure $G_C$ to the $s_1$-th child of the root and the $s_2$-th child of that node, using $\code{subColumn}$ queries, i.e., to the $\cC_1$ and $\cC_2$ columns that contain the query point. 

The $\cC_2$-column is stored as a column permutation. Within this we can identify, using  the column offset $s_3$, the cell index $c_2$ and within-cell vertical rank $v$. Recall that this means that the entry $(\tau_i,i)$ is in the $c_2$-th non-zero $2$-cell in its $\cC_2$-column and that this cell has $v$ entries below $(\tau_i,i)$.

Using the $\code{rowRank}$ and $\code{parentCell}$ queries on the $\cC_1$-column with the cell index $c_2$ we just found, and indicating that our $\cC_2$-column is the $s_2$-th of its parent, we obtain three values. These are $c'_2$, indicating that the $2$-cell where the query point falls is the $c'_2$-th non-zero cell in its $\cR_2$-row; $r_2$, meaning that this row is the $r_2$-th within its parent $\cR_1$-row, and $c_1$, indicating that the $1$-cell where the query point falls is the $c_1$-th non-zero cell in its $\cC_1$-column.  
Using $\code{rowRank}$ on the root, we can now determine that the $1$-cell containing the query point is in the $r_1$-th $\cR_1$-row. 

Using the row indices $r_1$ and $r_2$ we navigate the row tree $T_R$ to find the leaf corresponding to the $\cR_2$-row containing the queried element. By querying its global leaf rank we identify the $\cR_2$-row $r$ where the query falls.
Querying $I_R$ we compute the exact vertical offset $r_i$ of the row $r$ (i.e., the starting coordinate of this row within the matrix). 

It remains to compute the exact vertical offset within the row. For this we navigate $G_R$ using $r_1,r_2$, to find the row permutation containing the query point. From this row permutation, using the vertical rank $v$ and the horizontal index $c_2'$ of the $2$-cell among non-zero cells, we obtain the exact vertical offset $r_3$ within the row.

Our global vertical offset (and the answer to the query $\tau(i)$) is now obtained by adding $r_3$ to $r_i$.
As all steps take $\fO(1)$ time, the overall cost of rank (as well as unrank) queries is $\fO(1)$, independent of~$\pi$. The bound of $2n\lg{s_{\pi}} + o(n)$ on the space usage follows from the description of the components.

\paragraph{Other operations.} We briefly illustrate how other $\fO(1)$-time operations can be implemented on top of the main data structure with little overhead. We omit describing several symmetric (rotated) variants of these operations, as they can be implemented very similarly.

\medskip

We first discuss \textbf{range minimum}. As mentioned in \S\,\ref{sec:prelim}, with $2n + o(n)$ overhead, we can support range minimum queries over the permutation $\tau$ (and similarly, over $\tau^{-1}$). We show that this can also be achieved with sublinear overhead. Most of the steps are similar to those of \emph{rank} queries, we thus focus only on the necessary extensions. 

On each $\cC_2$-column we support queries $\code{rm}(a,b)$ where $a,b$ are relative column offsets within the 
$\cC_2$-column and the query returns the offset $k \in [a,b]$ of the minimum entry within the interval. As such queries depend only on the permutation within the $\cC_2$-column, they can be stored only once for all occurring (gridded) permutations, similarly to the $\code{colPerm(k)}$ implementation described in part C. Storing the answer to each possible query for a column of width $w$ requires $w^2 \cdot \lceil \lg{w} \rceil$ bits, for $w \in \fO(\sqrt{\lg{n}})$. Again, as there are at most $\fO(s_{\pi}^w \cdot w^{10c_{\pi}})$ such structures, the total space requirement is $o(n)$. 
Additionally, we build a range-minimum data structure over the minima of each $\cC_2$-column. Given two $\cC_2$-column indices $i_a$ and $i_b$, the structure returns the index $i_k \in (i_a,i_b)$ of the $\cC_2$ column with smallest minimum. The overhead of this structure is $2m_2 + o(m_2) \subseteq o(n)$.

Consider a query $\code{rm}(a,b)$. We first identify the $\cC_2$-columns where $a$ and $b$ fall. If $a$ and $b$ fall into the same $\cC_2$-column, we can directly answer the query with a $\code{rm}$-query within the column. (The answer is an offset within the column, but we can easily translate this into a global column index.)

If $a$ and $b$ fall into different $\cC_2$-columns, say $i_a$, $i_b$, then we perform four range-minimum queries: (1) in column $i_a$ from the position of $a$ to the end of the column, (2) in column $i_b$ from the beginning of the column to $b$, (3) on the column-minima for columns with index between $i_a$ and $i_b$, to identify column $i_k$ with the smallest minimum, and finally (4) on column $i_k$ to identify the actual minimum entry in this column. By using rank queries, we obtain the actual values of the entries returned in (1), (2), and (4), and return the index of the smallest among these. 

The space requirement for the additional structures we created is $o(n)$. Note that other types of queries over columns (or rows) can be similarly handled. 

\medskip

We next describe \textbf{next smaller} queries. Here, given input $i$, we wish to return the smallest $j>i$ for which $\tau_j < \tau_i$. 

First, given $i$, we identify the $\cC_2$-column $C$ and $\cR_2$-row $R$ where $(\tau_i,i)$ falls. We first verify whether there is a next smaller element in $C$; if there is, then that is the answer to the original query and we are done. 
As such queries depend only on the column permutation of a $\cC_2$-column, we can precompute the answers to all possible queries and store them similarly to the $\code{colPerm(k)}$ queries described in part C, in the table of column-permutations. We obtain the answer as an offset within $C$, and again, we can translate this into a global column index easily. The overhead is $o(n)$. 

Now we issue a similar query in $R$. We obtain the answer as an offset within $R$, which we resolve into a global column index similarly to \emph{unrank} queries. The obtained entry $(\tau_x,x)$, if exists, is a candidate for the overall next smaller.

Next, we look within the $\cC_2$-columns that are to the right of the $2$-cell of $(\tau_i,i)$, intersecting the $1$-cell of $(\tau_i,i)$. We look for the leftmost element in these columns that is ``above'' the $2$-cell of $(\tau_i,i)$ (recall that being above means having smaller $\tau$-value). We identify such an element by its column index within the $1$-cell, thus by $\fO(\lg\lg{n})$ bits. As the query depends only on the $2$-cell containing the input, we can precompute the answers for all non-zero $2$-cells, taking in total $\fO(n/\sqrt{\lg{n}} \cdot 10c_{\pi} \cdot \lg\lg{n}) \subseteq o(n)$ bits. The obtained index can again be resolved to a global column index easily. 

Symmetrically, we look for a candidate element within the $\cR_2$-rows that are above the $2$-cell of $(\tau_i,i)$, intersecting the $1$-cell of $(\tau_i,i)$. We look for the leftmost element in these rows that is to the right of the $2$-cell of $(\tau_i,i)$.

The only remaining location for the query answer is now fully to the right and above the $1$-cell of $(\tau_i,i)$, namely the leftmost entry in this area. Since there are only $m_1 \cdot 10c_\pi$ $1$-cells, we can precompute each such query, whose answer identifies the global column ($\fO(\lg{n})$ bits) that contains the solution. This takes $\fO(n/\lg^2{n} \cdot 10c_{\pi} \cdot \lg{n}) \subseteq o(n)$ bits. 

We now have (up to) four candidate points as the answer to the original \emph{next smaller} query. Each of these is feasible, so the overall answer is the leftmost among them. 

\emph{Remark:} We formulated our solution by decomposing the original query into queries over five areas: (1) the fine row of the input, (2) the fine column of the input, (3) the fine columns intersecting the coarse cell, to the right of the fine cell, (4) the fine rows intersecting the coarse cell, above the fine cell, and (5) the part of the matrix above and to the right of the coarse cell of the input. It is clear that these five areas together cover the area where the solution could lie. 
In certain applications it may be advantageous to decompose the query area into \emph{disjoint} parts -- in this way we can support various counting, and more general \emph{semigroup queries}, where the answer is formed by composing the answers over disjoint components, as is common in computational geometry. 
It is not hard to make the query areas disjoint, by restricting part (2) to the area strictly above the fine cell (using the gridding of the permutation), and (3) to the area strictly above the coarse cell containing the input. For the sake of simplicity, we have avoided this technicality in our description. 
We note that similar issues arise and are handled in more detail in \S\,\ref{sec4}.  


\paragraph{Building the data structure.}
We briefly argue that the time to construct the data structure of \Cref{thm1} is linear. The main task is that of constructing the decomposition; by Lemma~\ref{lem:decomp}, this requires $\fO(n\lg{c_\pi}) = \fO(n\lg{s_\pi})$ time. A key step, both in constructing the decomposition and in building the various components is that of \emph{sorting the input}; after this step, points of the input can be navigated both horizontally and vertically. By~\cite{Opler_sort}, sorting $\tau$ takes $\fO(n\lg{s_\pi})$ time. 
Constructing parts A and B clearly takes linear time. Storing the column and row permutations (parts C, E) again relies on efficient sorting. This takes $\fO(w\lg{s_{\pi}})$ time for a column (row) of width (height) $w$, adding to $\fO(n\lg{s_{\pi}})$ overall. Tabulating the permutations by size, isomorphism, and gridding, identifying and indexing the non-zero cells, and collecting the answers for each query take time linear in the representation, using straightforward (but slightly tedious) data structuring. The construction of the recursive row and column structures (parts D, E) and precomputation of \code{rowRank}, \code{colRank}, \code{parentCell}, \code{subColumn}, \code{subRow} can likewise be achieved by straightforward navigation of the input and its balanced decomposition structure; we omit a more detailed description.

\section{Extension to geometric queries}\label{sec4}

In this section, we extend the data structure to geometric queries, proving Theorem~\ref{thm3}. We will only describe the implementation of \emph{rectangle range counting} queries, as the changes necessary to implement \emph{range minimum} or other rectangle-based queries will be obvious.

We modify the original data structure of \Cref{thm1} in two major ways.
First, we extend the row and column permutations \code{rowPerm} and \code{colPerm} with precomputed geometric queries.
Second, we build a new hierarchy of $\fO(\lg\lg n)$ divisions with row and column sizes decreasing at a polynomial rate, alongside the original, two-level hierarchy.

Throughout the section, $(\cR_1, \cC_1)$ and $(\cR_2, \cC_2)$ refer to the two non-trivial divisions of the data structure in \Cref{thm1}.
We now define the sizes of divisions in the new hierarchy.
Let $w_0, w_1, \dots$ be a sequence where $w_0 = n$ and $w_i = \lfloor w_{i-1}^{5/6} \rfloor$, for $i \geq 1$, and let $\ell$ be the smallest integer such that $n/w_\ell \ge \lfloor n/\sqrt{\lg n}\rfloor$; observe that $\ell \in \Theta(\lg\lg n)$.  

We compute $\ell$ levels of the balanced decomposition (\Cref{lem:decomp}) for parameters $1 < m_1 < \dots < m_\ell < n$ where $m_i = \lfloor n/w_i\rfloor$ for $i < \ell$ and $m_\ell = \lfloor n/\sqrt{\lg n} \rfloor$.
Let us denote the obtained divisions by $(\cR'_1, \cC'_1), \dots , (\cR'_\ell, \cC'_\ell)$.
We have made the specific choice for $m_\ell$ in order to guarantee that $(\cR'_\ell, \cC'_\ell)$ is exactly the same as $(\cR_2, \cC_2)$.
(Since the algorithm of \Cref{lem:decomp} is deterministic, it will always output on the same input, the same division of a given size.) 
We additionally let $(\cR'_0, \cC'_0)$ and $(\cR'_{\ell+1}, \cC'_{\ell+1})$ denote the two trivial divisions where $|\cR'_0| = |\cC'_0| = 1$ and $|\cR'_{\ell+1}| = |\cC'_{\ell+1}| = n$.

The crucial property of this hierarchy of divisions is that the number of subcolumns of each column is relatively small compared to its width: 
each column in~$\cC'_i$ contains at most $40 \cdot m_{i+1}/m_i \in \fO(w_i^{1/6})$ columns of~$\cC'_{i+1}$ by Lemma~\ref{lem:decomp}(e), and the same holds analogously for rows.

\begin{figure}
\centering\includegraphics[scale=0.21]{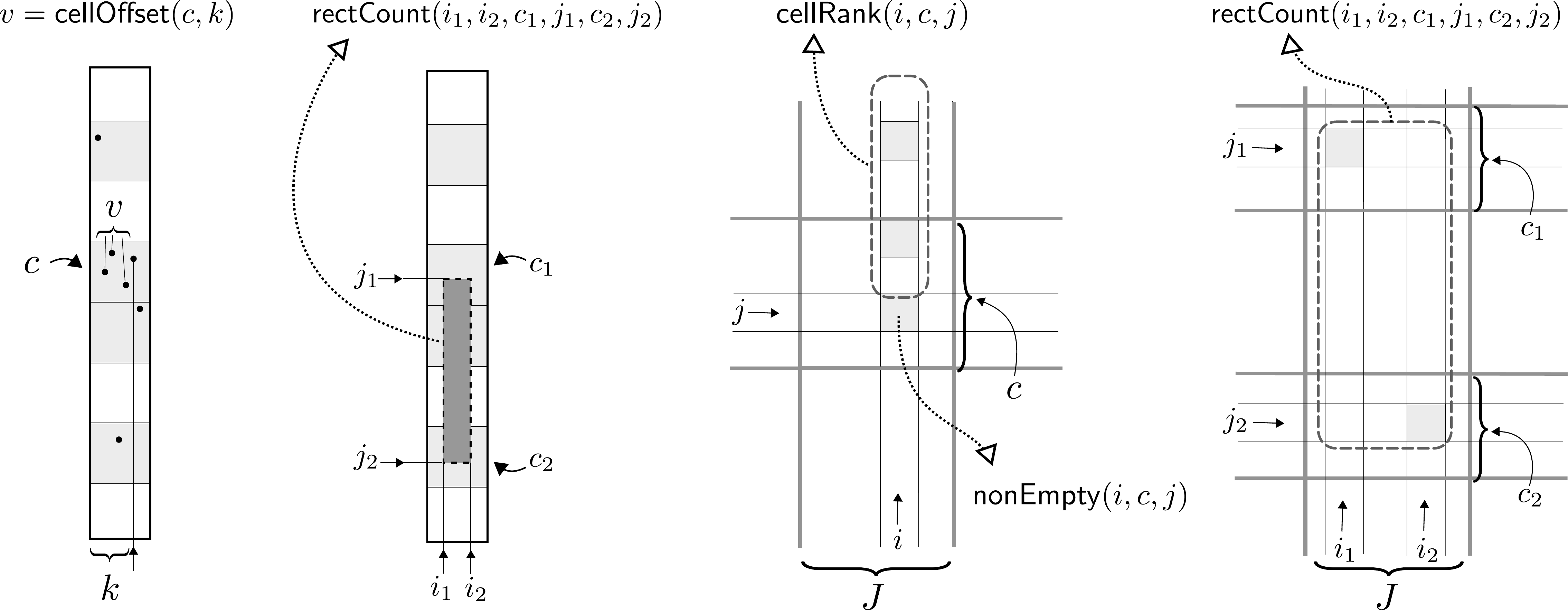}
		\caption{From left to right: (i) \code{cellOffset} query in $\cC_2$-column: for absolute offset $k$ and non-zero cell index $c$, we obtain the number of points $v$ left of the query; (ii) \code{rectCount} query in $\cC_2$-column: returns number of points in rectangle with corners identified by horizontal offsets $i_1,i_2$, and vertical offsets $j_1,j_2$ within non-zero cells $c_1$ and $c_2$; (iii) \code{cellRank} and \code{nonEmpty} queries for a column $J$ (at arbitrary level), w.r.\ to cell in subcolumn $i$ of $J$ and subrow $j$ within non-zero cell $c$ of $J$; \code{cellRank} returns number of non-zero cells in marked area; (iv) \code{rectCount} query for a column $J$ (at arbitrary level), $i_1,i_2$ now refer to subcolumns within $J$ and $j_1,j_2$ refer to subrows within cells $c_1, c_2$ of $J$. Returns number of points in rectangle.}\label{fig6}
        \end{figure}

We next describe the new or modified components of the data structure, referring to Figure~\ref{fig6} for an illustration.

\paragraph{C'. Extension of column and row permutations.}
We extend the column and row permutations with two additional queries. 
This will not affect the leading term in the space complexity, as this data is only stored in the precomputed tables, of sublinear total size. 

Let $J$ be a $\cC_2$-column (alternatively $\cC'_\ell$-column).
We support new queries \code{rectCount} and \code{cellOffset}, defined as follows:
\begin{itemize}
\item $\code{rectCount}(i_1, i_2, c_1, j_1, c_2, j_2)$: values $i_1, i_2$ are column offsets within~$J$, $c_1, c_2$ are (non-zero) cell indices, $j_1, j_2$ are relative row ranks within cells. The query returns the number of points in the column permutation, within the rectangle (including its boundaries) spanned by the $i_1$-th and $i_2$-th columns and the $j_1$-th row within the $c_1$-th cell and $j_2$-th row within the $c_2$-th cell.
\item $\code{cellOffset}(k, c)$: value $k$ is a column offset within column~$J$ and $c$ is a cell index. The query returns the number of points contained in the first $k$ columns of~$J$ within the $c$-th cell.
\end{itemize}

We store the answers to all these queries precomputed in the tables.
The total number of such queries is $\fO(c^2_\pi (\sqrt{\lg n})^4) = \fO_\pi(\lg^2 n)$ with each answer taking $\fO(\lg \lg n)$ bits.
The overall extra space needed in the precomputed tables is thus $\fO(s_\pi^w \cdot w^{10 c_\pi} \cdot c^2_\pi w^4 \lg w) \subseteq o(n)$, where we used $w = \fO(\sqrt{\lg n})$. 

For row permutations, a query $\code{cellOffset}(k,c)$ outputs the number of points in the $c$-th non-zero cell contained in the first $k$ subrows and a query $\code{rectCount}(i_1,i_2,c_1,j_1,c_2,j_2)$ returns the number of points within the rectangle spanned by the $i_1$-th and $i_2$-th rows and the $j_1$-th column within the $c_1$-th cell and the $j_2$-th column within the $c_2$-th cell.  

\paragraph{F. Dense column and row trees $T'_C, T'_R$.}
These trees are exact analogues of the trees $T_C$ and $T_R$ (part A), for the new, larger hierarchy of divisions,
supporting the same navigational queries as $T_C$ and $T_R$.
Again, these take a linear number of bits in the  total number of nodes, using the succinct tree implementation.
Therefore, the overall space requirement is $\fO(1 + \sum_{i=1}^\ell m_i) \subseteq \fO(\ell \cdot n/ \sqrt{\lg n}) \subseteq o(n)$ where the first inclusion holds since $m_i = \lfloor n/w_i \rfloor \leq \sqrt{n/\lg n}$ for every $i \in [\ell]$.

\paragraph{G. Dense recursive column and row structures $G'_C, G'_R$.}
Using the sequence of divisions, we build a data structure $G'_C$ similar to $G_C$ in part D of the original structure. (As $G'_R$ is entirely symmetric, we omit its description.)
Again, the root corresponds to the interval~$[n]$ (the single column of~$\cC'_0$), level-1 nodes correspond to the $m_1$ columns in~$\cC'_1$ and in general, level-$i$ nodes correspond to the $m_i$ columns in~$\cC'_i$ with the children of a node being its respective subcolumns within~$\cC'_{i+1}$.
We build $\ell$ levels but we keep the leaves on the $\ell$-th level without any stored information since we deal with the rows and columns of the division $(\cR'_\ell, \cC'_\ell) = (\cR_2, \cC_2)$ separately.
We maintain the same recursive representation in memory where each node consists of some local information followed by offsets into the final part, a concatenation of data structures of its children.
The tree is now significantly deeper, with $\ell \in \Theta(\lg\lg n)$ layers but importantly, its branching factor is smaller 
which allows for storing more information in each node.

For $k \in [\ell]$, a \emph{$k$-cell} is a submatrix $M[I,J]$ where $J$ is a $\cC'_k$ column and $I$ is a $\cR'_k$-row.
Let us consider a level-$k$ node that represents a $\cC'_k$-column~$J$.
We first describe a coordinate system we use to access the $(k+1)$-cells within $J$.
On the horizontal axis, we index by an explicit offset~$i$ (rank) among the $\cC'_{k+1}$-subcolumns of~$J$.
On the vertical axis, we describe the rows by a pair $(c, j)$ where $c$ is the rank of a non-zero $k$-cell within~$J$ and $j$ is an explicit offset (rank) among the $\cR'_{k+1}$-rows that intersect this $k$-cell.
Therefore, a particular $(k+1)$-cell within~$I$ can be uniquely identified by a triple $(i, c, j)$ and we say that it lies at coordinates $(i, c, j)$.

The node implements three novel methods (see Figure~\ref{fig6}): \code{nonEmpty}, \code{cellRank} and \code{rectCount}.
\begin{itemize}
\item $\code{nonEmpty}(i, c, j)$ indicates if the $(k+1)$-cell at coordinates $(i, c, j)$ is non-empty.
\item $\code{cellRank}(i, c, j)$ returns the number of non-empty $(k+1)$-cells above the $(k+1)$-cell at coordinates $(i, c, j)$ in the same $\cC'_{k+1}$-column, i.e., the number of non-empty $(k+1)$-cells at coordinates $(i, c, j')$ for $j' < j$ and at coordinates $(i, c', j'')$ for $c' < c$ and arbitrary $j''$.
We can, moreover, input $j = 0$ ($j = \infty$ resp.) to get the number of non-empty $(k+1)$-cells in the $i$-th subcolumn contained in the first $c-1$ ($c$ resp.) non-empty $k$-cells.
\item $\code{rectCount}(i_1, i_2, c_1, j_1, c_2, j_2)$ returns the number of points contained in the rectangle with opposite corners formed by the $(k+1)$-cells at coordinates $(i_1, c_1, j_1)$ and $(i_2, c_2, j_2)$ (inclusively).
As in \code{cellRank}, we can also set either of $j_1, j_2$ to $0$ or $\infty$.
By setting $j_1 = 0$, we include all rows in the $c_1$-th non-empty $k$-cell whereas setting $j_1=\infty$ includes only the rows strictly below the $c_1$-th $k$-cell.
Similarly, setting $j_2=0$ excludes all rows in the $c_2$-th non-empty $k$-cell whereas $j_2 = \infty$ includes all of them in the query; note that we assume $c_1<c_2$.
\end{itemize}

For simplicity, in all three cases we keep the signature of methods the same even for the root of~$G'_C$.
However, in that case the cell ranks ($c$, $c_1$ or $c_2$) must equal 1 since we consider the whole matrix~$M$ to be a single 0-cell.

We store all possible invocations of all three methods precomputed, available for constant-time retrieval.
Additionally, each node supports navigational $\code{subColumn}$ queries exactly as before.

Let us now bound the total size.
First, we prove a crude bound that allows us to bound the number of bits needed for the $\code{subColumn}$ offsets.
By reverse induction on $i$, we prove that each level-$i$ node fits into $w_i^2$ bits and in particular, the offsets needed for $\code{subColumn}$ therein fit in $\fO(\lg w_i)$ bits each.
The leaves (level-$\ell$ nodes) do not store any information, so the claim holds vacuously.
A general level-$i$ node has $\fO(w_i^{1/6})$ children, each taking up at most $w_{i+1}^2$ bits by induction.
Plugging in $w_{i+1} \le  w_i^{5/6}$, we get that the concatenation of the children's representations takes at most $\fO(w_i^{1/6} \cdot w_i^{10/6}) = \fO(w_i^{11/6})$ bits which is less than $\frac{1}{3}w_i^2$ for large enough~$n$ since $w_i \ge \sqrt{\lg n}$.
Thus, it suffices to reserve $2 \lg w_i$ bits for each $\code{subColumn}$ offset for a total of $\fO(w_i^{1/6} \lg w_i)$ bits which is, again, less than  $\frac{1}{3}w_i^2$.
Finally, the precomputed \code{nonEmpty}, \code{cellRank} and \code{rectCount} queries take $\fO_\pi\big(\big(w_i^{1/6}\big)^4  \lg w_i \big)\in \fO_\pi(w_i^{2/3}  \lg w_i)$ bits of space, again less than $\frac{1}{3}w_i^2$ for large enough~$n$.
Summing together, we get that level-$i$ nodes take at most $w_i^2$ bits in total, completing the induction. 

Now, we do a more fine-grained global analysis of the overall space complexity.
Observe that the final structure is simply a concatenation of the local data (\code{subColumn} offsets and precomputed queries) for each node as encountered during an in-order traversal of the tree.
Therefore, it suffices to bound the total space taken by the local data of each node.
We do this by levels.
On the $i$-th level, there are $m_i$ nodes and each needs $\fO(w_i^{1/6} \cdot \lg w_i)$ bits to store \code{subColumn} offsets and $\fO_\pi(w_i^{2/3}  \lg w_i)$ bits to store the precomputed queries.
Together, this makes $\fO_\pi(w_i^{2/3}\lg w_i)$ bits per level-$i$ node for a total of  $\fO_\pi(m_i \cdot w_i^{2/3} \lg w_i) = \fO_\pi(n \cdot w_i^{-1/3} \lg w_i)$ over the whole $i$-th level. 
Summing over all $\ell$ levels, we obtain an overall bound of $\fO_\pi(n \cdot \sum_{i=1}^\ell w_i^{-1/3} \lg w_i)$ bits.
Finally, plugging in $w_i \ge \sqrt{\lg n}$, we obtain the bound of $\fO_\pi(n \cdot \ell \cdot \lg^{-1/6} n \cdot \lg \lg n) \subseteq o(n)$ bits.

\subsection{Implementation of geometric queries}
Let us now describe the implementation of a rectangular counting query, that asks for the number of points within a rectangle $R$.
In the following, we generally use the four compass directions -- north, west, south, east -- referring to the four sides of the rectangle~$R$ -- top, left, bottom, right.

Given a rectangle $R = [x^N, x^S]\times[x^W, x^E]$, 
we first identify the location of its corners within the hierarchy of divisions.
This is done similarly as in the rank/unrank queries.
We find using \code{rank} and \code{select} queries over $I_R$ the indices $r^S, r^N$ of the $\cR'_\ell$-rows that contain $x^S, x^N$ and the relative offsets $r^S_{\ell+1}, r^N_{\ell+1}$ within these rows.
Next, we use $T'_R$ to look up the relative ranks $r^S_1, \dots, r^S_\ell$ and $r^N_1, \dots, r^N_\ell$ where $r^S_i$ ($r^N_i$ resp.) is the relative rank of the $\cR'_i$-row containing~$x^S$ ($x^N$ resp.) within its parent $\cR'_{i-1}$-row.
Analogously, we obtain from $I_C$ the indices $s^W, s^E$ of the $\cC'_\ell$-columns that contain $x^W, x^E$ and the relative offsets $s^W_{\ell+1}, s^E_{\ell+1}$ within these columns.
We use $T'_C$ to look up the relative ranks $s^W_1, \dots, s^W_\ell$ and $s^E_1, \dots, s^E_\ell$ where $s^W_i$ ($s^E_i$ resp.) is the relative rank of the $\cC'_i$-column containing~$x^W$ ($x^E$ resp.) within its parent $\cC'_{i-1}$-column.
Additionally, we denote by $v^\alpha_i$ for $\alpha\in\{S,N,W,E\}$ the level-$i$ node in $G'_C$ or $G'_R$ that contains the respective boundary of~$R$.
This phase performs constantly many queries to $I_R, I_C$ and two traversals of $T'_C, T'_R$ each, it thus takes $\fO(\lg\lg n)$ time.

\begin{figure}
\centering\includegraphics[scale=0.6]{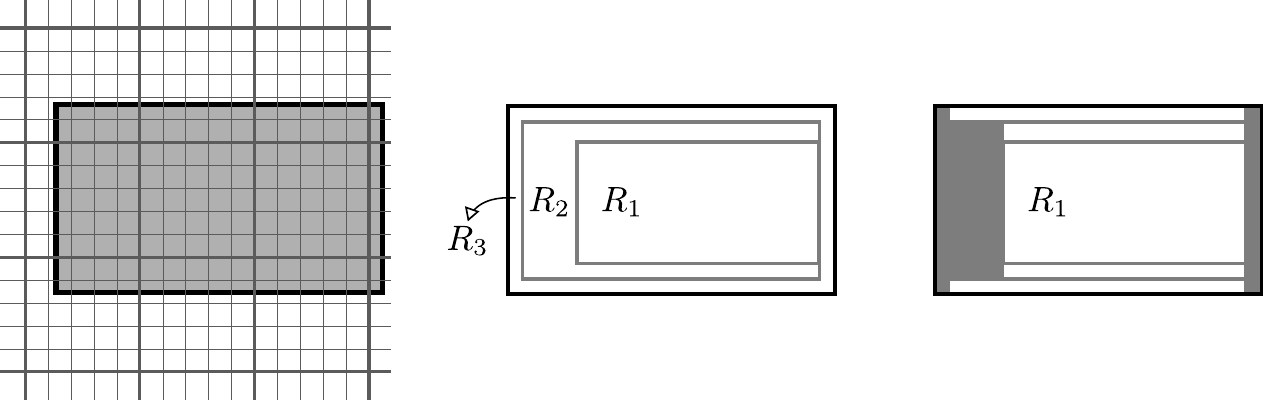}
		\caption{(i) Query rectangle $R$, with two levels of the decomposition shown. (ii) Decomposition of $R$ into layers $R_1, R_2, R_3$. (iii) West- and east frame shown gray, north and south frame empty (above and below $R_1$ region).}\label{fig7}
        \end{figure}

On a conceptual level, the answer is composed from precomputed \code{rectCount} counts similarly to the \emph{next smaller} queries in \S\,\ref{secthm1}.
We split the rectangle inductively into layers, where $R_1$ consists of all $1$-cells fully contained within~$R$, and for $i > 1$, the layer $R_i$ consists of the $i$-cells fully contained within $R \setminus (R_1 \cup \dots \cup R_{i-1})$.
Note that there are $\ell+1$ layers in total and we refer to $R_i$ as the \emph{$i$-layer} of~$R$.
Let $i_{\min}$ be the smallest integer such that there is an $i$-cell fully contained in~$R$.
By definition, $R_i = \emptyset$ for every $i < i_{\min}$.
Observe that $R_{i_{\min}}$ is a rectangle while $R_i$ for $i > i_{\min}$ is a set difference of two nested rectangles; see Figure~\ref{fig7}. 

Moreover for each $i > i_{\min}$, we split $R_i$ into four disjoint parts $R^E_i, R^S_i, R^W_i, R^N_i$, one per each side of the rectangle~$R$.
We define $R^W_i$ and $R^E_i$ as the intersection of $R_i$ with the leftmost and rightmost $\cC'_{i-1}$-columns that intersect~$R$.
In other words, $R^W_i$ ($R^E_i$ resp.) consist of all $i$-cells that are fully contained in~$R$ and lie within the leftmost (rightmost resp.) $\cC'_{i-1}$ column that intersects~$R$.
For $R^S_i$ and $R^N_i$, we similarly restrict $R_i$ to the bottommost and topmost $\cR'_{i-1}$-row intersecting~$R$.
However, we additionally exclude the $(i-1)$-cells that contain the corners of~$R$ since these are already included in $R^W_i$ and $R^E_i$.
We refer to the union $R^E_{i_{\min}+1} \cup \dots \cup R^E_\ell$ as the \emph{east frame} of~$R$.
The west, south and north frames are defined analogously.

\paragraph{Counting points in $R_1 \cup \dots \cup R_\ell$.}
Let us first assume, for the sake of simplicity, that all four corners of $R$ fall into \emph{different, empty} cells in the coarsest division~$(\cR'_1, \cC'_1)$.
In other words, each corner belongs to a different empty 1-cell and $i_{\min} = 1$.
In this case, we use a single $\code{rectCount}(s^W_1 + 1, s^E_1 - 1, 1, r^N_1 +1,1, r^S_1 - 1)$ query on the root of $G'_C$ to get the number of points in~$R_1$.
This is because $R_1$ consists exactly of the 1-cells strictly between the $s^W_1$-th and $s^E_1$-th columns, and $r^N_1$-th and $r^S_1$-th rows.
Next, we do two recursive traversals of $G'_C$ and two of $G'_R$, one per each side of the rectangle~$R$, to compute the contribution of east, west, north and south frames separately.
Since these are symmetric, we only describe here the case of the east frame of~$R$.

Fix $i >1$, and let $J$ be the $\cC'_{i-1}$-column that contains the right boundary of~$R$.
Horizontally, the rectangle $R^E_i$ spans exactly the first $s^E_i-1$ columns of~$J$ since the boundary falls into the $s^E_i$-th subcolumn.
We also know that $R^E_i$ is vertically aligned with the non-empty $(i-1)$-cells in~$J$ since the corners of~$R$ fall into empty 1-cells.
As we recurse into $G'_C$, we compute for each level-$i$ the values $c^E_{i,1}$ and $c^E_{i,2}$ such that $R^E_i$ spans vertically exactly the $c^E_{i,1}$-th up to the $c^E_{i,2}$-th $(i-1)$-cell.
For $i = 2$, we set
\begin{align*}
c^E_{2,1} &\gets G'_C.\code{root}.\code{cellRank}(s^E_1, 1, r^N_1) + 1\\
c^E_{2,2} &\gets G'_C.\code{root}.\code{cellRank}(s^E_1, 1, r^S_1)
\end{align*}
since we know the exact location of the northeast and southeast corners of~$R$.
For $i >2$, we compute $c^E_{i,1}, c^E_{i,2}$ from $c^E_{i-1,1}, c^E_{i-,2}$ at the node $v^E_{i-1}$ of~$G'_C$ as follows:
\begin{align*}
	c^E_{i,1} &\gets v^E_{i-1}.\code{cellRank}(s^E_i, c^E_{i-1,1}, 0) + 1\\
	c^E_{i,2} &\gets v^E_{i-1}.\code{cellRank}(s^E_i, c^E_{i-1,2}, \infty).
\end{align*}
This simply follows from the definition of $\code{cellRank}$ using the fact that $R^E_i$ is vertically aligned with the non-empty $(i-1)$-cells.
Having these, we can immediately get the number of points in~$R^E_i$ as
$v^E_{i-1}.\code{rectCount}(1,s^E_i-1,c^E_{i-1,1}, 0, c^E_{i-1,2}, \infty)$.

Repeating a symmetric procedure for the remaining three sides and adding the counts together, we obtain the desired number of points in $R_1 \cup \dots \cup R_\ell$ in time $\fO(\lg\lg n)$, since this amounts to four top-down traversals of $G'_C$ and $G'_R$ in total, with constantly many queries on each level.

\medskip

Now, let us describe how to deal with the two additional assumptions, the first being the emptiness of the 1-cells containing the corners of~$R$.
We still assume that all four corners of~$R$ fall into different 1-cells (and, thus, $i_{\min} = 1$) but they can now be non-empty.

We generalize the procedure counting points in the east frame to this scenario.
To that end, we additionally compute along the traversal for each level-$i$ two flags $b^E_{i,1}$ and $b^E_{i,2}$ where $b^E_{i,1} = \ttrue$ ($b^E_{i,2} = \ttrue$ resp.) if and only if the southeast (northeast resp.) corner of $R$ falls into a non-empty $i$-cell.
We also specify the semantics (and computation) of $c^E_{i,1}$ and $c^E_{i,2}$ in this case.
These now contain the minimum and maximum rank of non-empty $(i-1)$-cells within the respective $\cC'_{i-1}$-column that intersects~$R$.
Observe that the semantics agree in the previous restricted case of empty 1-cells.

For $i = 1$, the values $b^E_{1,1}$ and $b^E_{1,2}$ are obtained as
\begin{align*}
	b^E_{1,1} &\gets G'_C.\code{root}.\code{nonEmpty}(s^E_1, 1, r^N_1)\\
	b^E_{1,2} &\gets G'_C.\code{root}.\code{nonEmpty}(s^E_1, 1, r^S_1).
\end{align*}

For $i >1$, we describe the computation of $b^E_{i,1}$ only, as the computation of $b^E_{i,2}$ is analogous.
First, observe that if a corner of~$R$ falls into an empty $(i-1)$-cell then it necessarily falls into an empty $i$-cell.
Otherwise, the southeast corner lies in the $c^E_{i-1,1}$-th non-empty $(i-1)$-cell.
We set
\[b^E_{i,1} \gets \begin{cases}
\tfalse &\text{if $b^E_{i-1,1} = \tfalse$, and}\\
v^E_{i-1}.\code{nonEmpty}(s^E_i, c^E_{i-1,1}, r^N_i) &\text{otherwise,}
\end{cases}\]
where we use the fact that the relative vertical offset of the northeast corner within its $(i-1)$-cell is exactly equal to $r^N_i$, that is, the relative rank of the $\cR'_{i}$-row containing the north boundary of~$R$ within its $\cR'_{i-1}$ parent.
We similarly adapt the computation of $c^E_{i,1}$:
\[c^E_{i,1} \gets \begin{cases}
	v^E_{i-1}.\code{cellRank}(s^E_i, c^E_{i-1,1}, 0) + 1 &\text{if $b^E_{i-1,1} = \tfalse$, and}\\
	v^E_{i-1}.\code{cellRank}(s^E_i, c^E_{i-1,1}, r^N_i) + 1 &\text{otherwise.}
\end{cases}\]

Finally, we count the number of points in~$R^E_i$ as $v^E_{i-1}.\code{rectCount}(1,s^E_i-1,c^E_{i-1,1}, j_1, c^E_{i-1,2}, j_2)$ where
\[j_1 = \begin{cases}
0	&\text{if $b^E_{i-1,1} = \tfalse$,}\\
r^S_i+1 	&\text{otherwise.}
\end{cases} \qquad
j_2 = \begin{cases}
	\infty	&\text{if $b^E_{i-1,2} = \tfalse$,}\\
	r^N_i-1 	&\text{otherwise.}
\end{cases}
\]
The modified procedure still takes $\fO(1)$ time on each level for a total of~$\fO(\lg\lg n)$ time.

The call to \code{rectCount} is symmetric for the west frame of~$R$.
However, we need to to modify the counting for the south and north frames to avoid overcounting.
We only describe the difference for the north frame.
We are computing the values $c^N_{i,1}, c^N_{i,2}$ and $b^N_{i,1}, b^N_{i,2}$ such that $R^N_i$ intersects horizontally exactly the $c^N_{i,1}$-th up to $c^N_{i,2}$-th non-empty $i$-cells in the corresponding $\cC'_{i-1}$-row and $b^N_{i,1}, b^N_{i,2}$ capture whether the northheast and northwest corners of~$R$ fall into a non-empty $i$-cell.
These values are computed analogously to the computation for the east frame described above.
However, the final \code{rectCount} call changes to $v^N_{i-1}.\code{rectCount}(1,r^N_i-1,c^N_{i,1}, j_1, c^N_{i,2}, j_2)$ where
\[j_1 = \begin{cases}
	0	&\text{if $b^N_{i-1,1} = \tfalse$,}\\
	\infty 	&\text{otherwise.}
\end{cases} \qquad
j_2 = \begin{cases}
	\infty	&\text{if $b^N_{i-1,2} = \tfalse$,}\\
	0 	&\text{otherwise.}
\end{cases}
\]
This holds since the left and right boundaries of $R^N_i$ are always aligned with $\cC'_{i-1}$-columns and we only have to determine whether to include the  $c^N_{i,1}$-th and $c^N_{i,2}$-th cell depending on whether they contain a corner of~$R$.

\medskip
Finally, we need to remove the assumption that all corners of~$R$ fall in different 1-cells.
One possibility is that there exists $i$ such that the four corners fall into the same $(i-1)$-cell but in four different $i$-cells.
Observe that in that case we have $i_{\min} = i$ and the union $R_1 \cup \dots \cup R_{i-1}$ of first $i-1$ layers is empty.
We use the same top-down traversal, except we count points in $R_i$ using a four-sided query at the node $v^E_{i-1}$ and only start counting points in the four frames from level-$i$.
However, we compute the values $c^E_{i,\cdot}$ and $b^E_{i,\cdot}$ (and analogously for the other three sides) starting at the root exactly as before.

The second option is that there are different $i_1$ and $i_2$ such that the right and left boundaries of~$R$ fall within the same $\cC'_{i_1-1}$-column but in different $\cC'_{i_1}$-columns, and the bottom and top boundaries of~$R$ fall within the same $\cR'_{i_2-1}$-row but in different $\cR'_{i_2}$-rows.
Let us assume that $i_1 < i_2$, the other case being analogous.
In this case, $i_{\min} = i_2$ and the union $R_1 \cup \dots \cup R_{i_2-1}$ is empty since~$R$ cannot fully contain any $i'$-cell for $i' < i_2$.
If $i_2 = \ell + 1$, then $R_1 \cup \dots \cup R_\ell$ is empty and the query is fully resolved within the fine rows and columns in~$\fO(1)$ time as described below.
Otherwise, we proceed as before with counting points in $R_{i_2}$ at level-$(i_2-1)$ and starting counting points in the four frames from level-$i_2$.

\paragraph{Counting points in $R_{\ell+1}$.}
We consider two cases depending on whether all four corners of~$R$ lie in different $\ell$-cells, or the rectangle $R$ fits into a single $\cC'_{\ell}$-column (the case of a single $\cR'_\ell$-row is symmetric).

In the first case, we want to count the points in $R_{\ell+1}$ in each of the four frames separately using \code{rectCount} queries.
We again describe the computation for the east frame only.
We have previously computed the values $c^E_{\ell,1}, c^E_{\ell,2}$ and $b^E_{\ell,1}, b^E_{\ell,2}$ at level-$(\ell-1)$ in~$G'_C$.
If both $b^E_{\ell,1}$ and $b^E_{\ell,2}$ are $\tfalse$, we query $\code{rectCount}(1,s^E_{\ell+1},c^E_{\ell,1}, 0, c^E_{\ell,2}, \infty)$ on the finest column that contains the east boundary of~$R$, using the column offset $s^E_{\ell+1}$ that was obtained from~$I_C$ at the beginning of the query.
Otherwise, we cannot simply plug in the row offsets $r^N_{\ell+1}$ and $r^S_{\ell+1}$ for $j_1$ and $j_2$ since the finest columns are stored as permutations, potentially losing information about empty rows.

Nevertheless, the correct relative row offsets $j_1, j_2$ can be easily computed using \code{cellOffset} queries.
If $b^E_{\ell,1} = \tfalse$, we have $j_1 = 0$  and otherwise, the desired row offset~$j_1$ is given precisely by the query $\code{cellOffset}(r^N_{\ell+1}, c^N_{\ell,2})$ on the $\cR'_\ell$-row that contains the north boundary of~$R$.
Similarly, if $b^E_{\ell,1} = \tfalse$, we set $j_2 = \infty$ and otherwise, $j_2 = \code{cellOffset}(r^S_{\ell+1}, c^S_{\ell,2})$ on the $\cR'_\ell$-row that contains the south boundary of~$R$.

It remains to deal with the case when, without loss of generality, $R$ (and thus $R_{\ell+1}$) fits into a single $\cC'_\ell$-column.
However, in that case the cell ranks $c^N_{\ell,1}$ and $c^N_{\ell, 2}$ must be equal because the northwest and northeast corners fall into the same $\ell$-cell.
The same holds for $c^S_{\ell,1}$ and $c^S_{\ell, 2}$.
Therefore, we first compute the relative row offset~$j_1$ of the north boundary by querying $\code{cellOffset}(r^N_{\ell+1}, c^N_{\ell,2})$ on the $\cR'_\ell$-row containing the north boundary.
Similarly, the relative row offset~$j_2$ is obtained by querying $\code{cellOffset}(r^S_{\ell+1}, c^S_{\ell,2})$ on the $\cR'_\ell$-row with the south boundary.
The number of points in $R_{\ell+1}$ is then given by the query $\code{rectCount}(s^W_{\ell+1},s^E_{\ell+1},c^E_{\ell,1}, j_1, c^E_{\ell,2}, j_2)$.

\section{Bounded treewidth case}\label{sec5}

In this section, we prove Theorem~\ref{thm2}. A key concept we use, in the context of balanced decompositions, is that of a \emph{component}. 

\begin{figure}
\centering\includegraphics[scale=0.6]{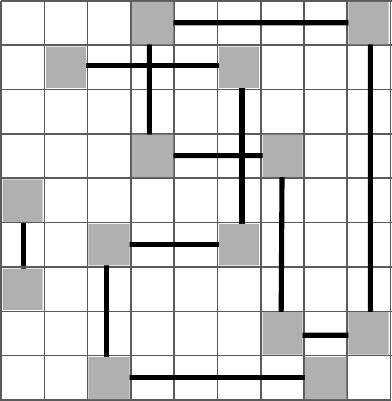}
		\caption{Three components of a division, with non-zero cells shaded gray, edges indicated with thick lines.}\label{fig8}
        \end{figure}

Given a division, we consider the graph whose vertices are the non-zero cells of the division and two vertices are connected by an edge exactly if the two corresponding cells are in the same column or row of the division. We refer to this graph as the graph of the division, and to its connected components as the \emph{components of the division}, see Figure~\ref{fig8}. The key observation is that if the treewidth of $\tau$ is bounded, then we can find a balanced decomposition similar to the one in Lemma~\ref{lem:decomp}, with the additional property that components of the division are of bounded size. This is captured in the following result, with proof in \S\,\ref{sec7}.

\begin{lemma}\label{lem:decomp_tw}
Let $\sigma$ be a permutation of length~$n$ and treewidth~$t$.
	
	For integer parameters $1 < m_1 < m_2 < \dots<  m_\ell < n$, there exist divisions $(\cR_1, \cC_1), \dots, (\cR_\ell, \cC_\ell)$ of~$M$ such that:
\begin{enumerate}[(a)]
	\item The division~$(\cR_i, \cC_i)$ is a coarsening of $(\cR_{i+1}, \cC_{i+1})$ for each $i \in [\ell-1]$.
	\item 
	$M(\cR_i, \cC_i)$ 
	has exactly $m_i$ rows and columns.\label[ipart]{item:bal-gridding-tw:num-rows}
	\item Each row in~$\cR_i$ and each column in $\cC_i$ has height, resp., width at most $\fO(t \cdot n/m_i)$.\label[ipart]{item:bal-gridding-tw:width-rows}
	\item Each component in~$(\cR_i, \cC_i)$ spans at most $\fO(t)$ rows and columns.
	\label[ipart]{item:bal-gridding-tw:points-per-row}
	\item Each row in~$\cR_i$ and each column in $\cC_i$ is obtained from merging at most $\fO(t \cdot m_{i+1}/m_i)$ rows of $\cR_{i+1}$, resp., columns of~$\cC_{i+1}$, 
	for $i \in [\ell-1]$. \label[ipart]{item:bal-gridding-tw:width-density}
\end{enumerate}
Moreover, the divisions $(\cR_1, \cC_1), \dots, (\cR_\ell, \cC_\ell)$ can be computed in time $\fO_t(n + \sum_{i=1}^{\ell}m_i)$.
\end{lemma}

Using Lemma~\ref{lem:decomp_tw}, the data structure for storing a permutation simplifies. Instead of the recursive row- and column- structures we stored earlier, we now build a single recursive structure based on the components.  
Assuming bounded treewidth, components are still sufficiently small to be stored in a global, precomputed table. 
A component-permutation allows converting, for the entries in the component, absolute column-information to row-information and vice versa.

We remark in passing that the notion of component and the decomposition in Lemma~\ref{lem:decomp_tw} relate to the concept of \emph{component twin-width}~\cite{BonnetKRT22}. Loosely speaking, we are computing a \emph{balanced} component twin-width decomposition, analogously to how  Lemma~\ref{lem:decomp} achieves a balanced twin-width decomposition.

\medskip

Again, we consider divisions $(\cR_1,\cC_1)$ and $(\cR_2,\cC_2)$ of the input, with the same parameters as in \S\,\ref{secthm1}, namely $m_1=n/{\lg^2{n}}$ and $m_2 = n/\sqrt{\lg{n}}$.
We invoke Lemma~\ref{lem:decomp_tw}, observing the new property that the components of both divisions are of size $O(t)$, where $t$ is the treewidth of $\tau$. In the following, assume that $t$ is constant.

The following parts of the data structure differ from the one in \S\,\ref{secthm1}. 

\paragraph{C'. Component permutations (replacing both column and row permutations).}

Each component of $(\cR_2, \cC_2)$ is stored as a gridded permutation. By \emph{gridded permutation} we now mean the permutation of points in a component, together with the horizontal and vertical partitioning, according to the different cells $(\cR_2, \cC_2)$ where they fall. (Contrast this with the column permutations in \S\,\ref{secthm1} that are only vertically split into cells, and row permutations that are only horizontally split.)

\begin{figure}
\centering\includegraphics[scale=0.6]{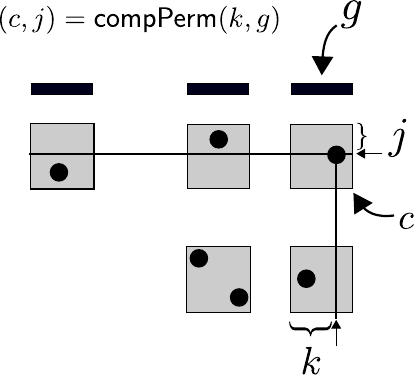}
		\caption{Given the (gridded) permutation of a component, we can identify, from the horizontal offset $k$ within gridding column $g$ the non-zero cell $c$ where the query falls and the absolute vertical offset $j$ within its row. Gray boxes indicate cells of the component, thick lines above indicate gridding columns. Dots other than the query are to be interpreted schematically as possibly multiple points.}\label{fig9}
        \end{figure}

The representation supports the query $\code{compPerm}(k,g) \rightarrow (c,j)$. The interpretation is that the entry in the $k$-th column within the $g$-th column of the gridding falls into the $c$-th non-zero cell with relative row offset~$j$; see Figure~\ref{fig9}. 
It also supports the query $\code{compPerm}^{-1}(k,g) \to (c,j)$, where the entry in the $k$-th row within the $g$-th row of the gridding falls into the $c$-th non-zero cell of the component with relative column offset~$j$. Non-zero cells within a component are indexed in an arbitrary canonical way, say first left to right, then top to bottom. 

The domain of both queries is $\fO(\sqrt{\lg{n}}) \times \fO(t)$ and their range is $\fO(t) \times \fO(\sqrt{\lg{n}})$. Thus, the precomputed answers for all such queries for a component gridded permutation take $\fO(\sqrt{\lg{n}} \cdot t \cdot (\lg\lg{n} + \lg{t}))$ bits.

Suppose that the total width of a component is $w$. Using an upper bound of $\fO(\sqrt{\lg{n}})$ on the column width, and the fact that a component consists of at most $O(t)$ cells, the number of possible gridded permutations that arise in a component is at most $s_\pi^w \cdot \sqrt{\lg{n}}^{\fO(t)}$, which means that an index to the component-permutation can be implemented in $w\lg{s_{\pi}} + \fO(\lg\lg{n})$ bits, which, over all components adds up to $n\lg{s_\pi} + o(n)$ bits. (Notice that no two components can overlap in a column of $\cC_2$, as otherwise they would merge, and that the number of components is at most $m_2 = n/\sqrt{\lg{n}}$.)
In case of a supermultiplicative subclass $\cC$ of $\pi$-avoiding permutations the total bound changes accordingly to $\lg|\cC_n| + o(n)$ bits.

Storing all the precomputed queries for at most $s_\pi^w \cdot \sqrt{\lg{n}}^{\fO(t)}$ distinct gridded permutations requires $o(n)$ bits overall, by a similar calculation as earlier. 

\paragraph{D'. Recursive component structure $G$ (replacing both $G_C$ and $G_R$).}
This is a three-level tree similar to $G_C$ and $G_R$.

We already discussed the storage of components of $(\cR_2,\cC_2)$. We similarly define components of $(\cR_1,\cC_1)$. We say that a component of $(\cR_2,\cC_2)$ is a subcomponent of a component of $(\cR_1,\cC_1)$, if they intersect in any entry of the input. Notice that this immediately means that the larger component fully contains the smaller (if the two components overlap on some entry, then every cell of the smaller must be contained in some cell of the larger).

The root of $G$ corresponds to the entire input, and level-1, resp., level-2 nodes correspond to $(\cR_1,\cC_1)$, resp., $(\cR_2,\cC_2)$ components, by containment. 

For level-1 nodes, the data structure supports the query $\code{subComponent}(k,g) \to (g',p)$ where the $k$-th subcolumn within the $g$-th column of the gridding falls into the $g'$-th column of the gridding of subcomponent~$p$.
Similarly, $\code{subComponent}^{-1}(k,g) \to (g',p)$ means that the $k$-th subrow within the $g$-th row of the gridding falls into the $g'$-th row of the gridding of subcomponent~$p$.
For the root the queries are similar, but with the second parameter $g$ missing (since the entire matrix can be viewed as a single cell/component where we need not provide gridding column or row). 

Additionally, the root supports $\code{rowRank}(k, c) \to r$ and $\code{colRank}(k, c) \to r$ implemented exactly as in \S\,\ref{secthm1}.
Level-1 nodes (components of $(\cR_1,\cC_1)$ support the query $\code{rowRank}(k, c) \to (c', r)$ and $\code{colRank}(k, c) \to (c', r)$ where $c$ and $c'$ now refer to the non-zero index of a cell in its component, but otherwise they are identical to the methods in \S\,\ref{secthm1}.

Level-2 nodes store a pointer to the component permutation structure that can be queried, as discussed in part C'.


The total space usage can be bounded almost identically to $G_C$ and $G_R$ in \S\,\ref{secthm1}, with components playing the role of columns (or rows), we therefore omit the detailed calculation. Notice that the number of components cannot be larger than the number of rows or columns in any division. 
In fact, now the number of non-zero cells in columns (and rows) is also at most $O(t)$. 

The saving of a factor two compared to Theorem~\ref{thm1} comes from the fact that we need not separately store row and column structures, but the single component structure suffices. The individual methods that we had to implement in duplicate for rows/columns all have sublinear contribution to the total space, the bottleneck is the table of component-permutations and its indexing, which is stored only once.  

\paragraph{Implementation of the queries.}

In case of \textbf{rank queries} (Figure~\ref{alg:rank-tw}), as before, we start by obtaining column indices and offsets from $I_C$ and $T_C$.
Then we navigate the component structure $G$ using \code{subComponent} queries, to obtain the component of $(\cR_1,\cC_1)$ and $(\cR_2,\cC_2)$ that contain the query point. 
Using the component of the fine division, we get via $\code{compPerm}$ the corresponding permutation, in which we resolve the query, obtaining the index of the non-zero cell in which the query falls and its vertical offset within its cell. 

Tracing up through $G$, we obtain the row ranks at the coarse level from the coarse level component, and at the root, using $\code{rowRank}$ queries. 

The endgame is exactly as in Theorem~\ref{thm1}; having obtained the row offset within the cell and relative row ranks within $\cR_1$ and $\cR_2$, we navigate $T_R$ and $I_R$ to compute the absolute row index which is the final output.  




The \textbf{unrank queries} can now be implemented similarly. We first obtain row indices and offsets from $I_R$ and $T_R$. Then we navigate the component structure to obtain the components containing the query point, this time with $\code{subComponent}^{-1}$ queries. Then we identify the non-zero cell in which the query falls, and its horizontal offset via $\code{compPerm}^{-1}$ queries. Going to the higher levels of~$G$, we find via \code{colRank} queries the column ranks at the coarse- and top levels. Finally, we navigate~$T_C$ and $I_C$ to compute the absolute column index, which is the output of the query.

\begin{figure}
	\small
	\textbf{Input:} $i$\\
	\textbf{Output:} $\tau(i)$
	\begin{algorithmic}
		\State $s \gets I_C.\srank(i)$ \Comment{Find column rank}
		\State $s_i \gets I_C.\sselect(s)$ \Comment{Column start index}
		
		\State $s_3 \gets i-s_i$ \Comment{Offset within column}
		\State $\mathit{col} \gets T_C.\code{leafSelect}(s)$ \Comment{Find relative column ranks}
		\State $s_2 \gets T_C.\code{childRank}(\mathit{col})$ 
		\State $s_1 \gets T_C.\code{childRank}(T_C.\code{parent}(\mathit{col}))$
		\State $\mathit{root} \gets G.\code{root}$ \Comment{Navigate component structure}
		\State $(g_1, \mathit{comp}_1) \gets \mathit{root}.\code{subComponent}(s_1)$ 
		\State $(g_2, \mathit{comp}_2) \gets \mathit{comp}_1.\code{subComponent}(s_2, g_1)$

		\State $(c'_2,r_3) \gets \mathit{comp}_2.\code{compPerm}(s_3,g_2)$ \Comment{Cell id and offset-within-cell read from component permutation}
		\State $(c'_1, r_2) \gets \mathit{comp}_1.\code{rowRank}( s_2,c'_2)$ \Comment{Parent cell id and relative row ranks}
		\State $r_1 \gets \mathit{root}.\code{rowRank}(s_1,c'_1)$ 
		\State $row \gets T_R.\code{child}(T_R.\code{child}(T_R.\code{root},r_1),r_2)$
		\Comment{Find row rank}
		\State $r \gets T_R.\code{leafRank}(\mathit{row})$
		\State $r_i \gets I_R.\srank(r)$ \Comment{Row start index}
		
		\textbf{return} $r_i +r_3$
	\end{algorithmic}
			\caption{Implementation of the rank query for bounded treewidth permutations.}\label{alg:rank-tw}
\end{figure}

\paragraph{Remark.}  Intuitively, the advantage of the component-based approach, compared to the earlier row/column-scheme (\S\,\ref{secthm1}), is that components fully span both rows and columns, thus, collapsing them to a permutation loses no information. To appreciate this difference, recall that in a column permutation, having identified a point in a cell $c$ with a \code{colPerm} query, we could only extract the relative rank $v$ of the point \emph{within its cell} (number of points below it) -- see Figure~\ref{fig_comp2} (left). 
This is because we did not have access to other non-zero cells in the same row as $c$ that may also contain points below the query, so we could not directly compute the rank of the query point within the entire row. (Put differently, this is exactly why we needed to access the row permutation as well.) In contrast, now \emph{all} non-zero cells in the same row or same column as $c$ are in the same component, and thus, all the necessary information to compute the exact vertical or horizontal rank is locally available.

We note that combining the design of this section with elements of the $O(\lg\lg{n})$-level hierarchy of \S\,\ref{sec4}, we could support similar geometric queries with sublinear space overhead, i.e., succinctly (assuming bounded treewidth); we omit the details of this extension. 



\section{Proof of the structural lemma}\label{sec6}
In this section we prove the balanced decomposition result (Lemma~\ref{lem:decomp}).

We first describe an algorithm that requires knowledge of the pattern~$\pi$ and in particular its Füredi-Hajnal limit~$c_\pi$.
Afterwards, we describe the generalization to unknown~$\pi$ borrowing ideas from the optimal pattern-avoiding-sorting algorithm~\cite{Opler_sort}.

\paragraph{Prior knowledge of~$c_\pi$.}
Let $d = 5c_\pi$, $\delta = 20$.
We say that a row $I$ in $M(\cR, \cC)$ is \emph{tall} if $|I| > \delta n/|\cR|$, and a column $J$ is \emph{wide} if $|J| > \delta n/|\cC|$.

We first give a high-level description of the algorithm and only afterwards we describe its efficient implementation in full detail.
We additionally define $m_0 = 1$ and $m_{\ell+1} = n$.
The algorithm iteratively computes a sequence of divisions $(\cR'_n, \cC'_n), \dots, (\cR'_1, \cC'_1)$ such that
\begin{enumerate}[(i)]
\item The division $(\cR'_{i}, \cC'_{i})$ is a coarsening of $(\cR'_{i+1}, \cC'_{i+1})$.\label[ipart]{item:rici:coarsening}
\item The division $(\cR'_{i}, \cC'_{i})$ has exactly $i$ rows and columns.\label[ipart]{item:rici:rows-cols}
\item Each row (resp.\ column) of $M(\cR'_i,\cC'_i)$ has 
height (resp.\ width) at most $2\delta \cdot n/i$.\label[ipart]{item:rici:width-height}
\item Each row and column in~$M(\cR'_i,\cC'_i)$ is obtained from merging at most $2 \delta \cdot m_{j+1}/m_j$ rows or columns of~$M(\cR'_{m_{j+1}}, \cC'_{m_{j+1}}) $, 
where $j$ is the largest integer such that $m_j \le i$.\label[ipart]{item:rici:non-dense}
\item Each row and column in~$M(\cR'_i,\cC'_i)$ 
contains at most $d$ non-zero cells. 
\label[ipart]{item:rici:sparse}
\end{enumerate}
Observe that we get the desired output by setting $(\cR_i, \cC_i) = (\cR'_{m_i}, \cC'_{m_i})$.
Thus, from now on we focus on computing the sequence $(\cR'_n, \cC'_n), \dots, (\cR'_1, \cC'_1)$. 

Initially, we set $(\cR'_n, \cC'_n)$ to be the trivial (finest) division where each row (resp.\ column) contains exactly one row (resp.\ column) of the input matrix~$M$.
We then run $\ell+1$ phases with the $j$-th phase generating divisions $(\cR'_i, \cC'_i)$ for $i \in \{m_j, \dots, m_{j+1}-1\}$.
We say that a row in the $j$-th phase is \emph{dense} if it contains more than $\delta \cdot m_{j+1}/m_j$ rows of the division $(\cR'_{m_{j+1}}, \cC'_{m_{j+1}})$, a dense column is defined analogously.

The algorithm generates $(\cR'_{i}, \cC'_{i})$ from $(\cR'_{i+1}, \cC'_{i+1})$ in the following way.
First, it finds an arbitrary pair of adjacent rows $R_1, R_2 \in \cR'_{i+1}$ such that $R_1$ and $R_2$ are both neither tall nor dense, and the row $R_1 \cup R_2$ contains at most $d$ non-empty cells with respect to~$\cC'_{i+1}$.
Similarly, it finds an arbitrary pair of adjacent columns $C_1, C_2 \in \cC'_{i+1}$ such that $C_1$ and $C_2$ are both neither wide nor dense, and the column $C_1 \cup C_2$ contains at most $d$ non-empty cells with respect to~$\cR'_{i+1}$.
Afterwards, the division $(\cR'_{i}, \cC'_{i})$ is obtained from $(\cR'_{i+1}, \cC'_{i+1})$ by replacing~$R_1$, $R_2$ and $C_1$, $C_2$ with their respective unions.

\paragraph{Correctness.}
For now, let us assume that the algorithm is able to generate all $n$ divisions.
In that case, we show that $(\cR'_i, \cC'_i)$ satisfies all the properties \ref{item:rici:coarsening}--\ref{item:rici:sparse} by reverse induction on~$i$.
Clearly, all of them hold for the initial division $(\cR'_n, \cC'_n)$.
Now, assume that $i < n$.
The properties \ref{item:rici:coarsening} and \ref{item:rici:rows-cols} follow directly from the algorithm.
Towards showing~\ref{item:rici:width-height}, assume that there exists a row~$R$ in $\cR'_i$ taller than $2\delta \cdot n/i$.
It must have been obtained by merging two non-tall rows in some step $k$ for $k > i$, i.e., each has height at most $\delta \cdot n/k$.
But then the height of~$R$ is at most $2\delta \cdot n/k \le 2\delta\cdot n/i$.
Symmetrically, \ref{item:rici:width-height} holds also for columns.
The situation of~\ref{item:rici:non-dense} is similar.
Any row~$R$ in~$(\cR'_i, \cC'_i)$ that contains more than one row (column) of the division $(\cR'_{m_{j+1}}, \cC'_{m_{j+1}})$ was obtained by merging two non-dense rows in $k$th step where $j+1 \ge k > i$.
A non-dense row in $k$th step contains at most $\delta \cdot m_{j+1}/m_j$ rows of  $(\cR'_{m_{j+1}}, \cC'_{m_{j+1}})$ and thus, we get that $R$ contains at most $2\delta \cdot m_{j+1}/m_j$ rows of~$\cR'_{m_{j+1}}$.
Analogous arguments show the same inequality for columns.
Finally, notice that merging pairs of adjacent rows cannot increase the number of non-empty cells in any fixed column and vice versa.
Moreover, the algorithm merges adjacent rows (columns) only if their union contains at most $d$ non-empty cells. 
Therefore, the algorithm never creates a row or column with more than $d$~non-empty cells and~\ref{item:rici:sparse} holds.

Assume for a contradiction that the algorithm was not able to construct the division~$(\cR'_i, \cC'_i)$ from $(\cR'_{i+1}, \cC'_{i+1})$ for some $i \in [n-1]$.
Let $j$ be the index of the corresponding phase, i.e., the largest integer such that $m_j \le i$.
We only consider the case when there was no suitable pair of adjacent rows~$R_1$ and $R_2$ as the case of columns is analogous.

 
Let us denote by~$T$ and $D$ the number of tall rows and dense rows, respectively.
First observe that $T < \frac{i+1}{\delta}$ since each tall row has size strictly larger than~$\delta \cdot \frac{n}{i+1}$ and thus, $T \le \frac{i}{\delta}$ as $\delta$ is an integer.
Moreover, the number of dense rows can be bounded as 
\[D < \frac{m_{j+1}}{\delta \frac{m_{j+1}}{m_j}} = \frac{m_j}{\delta} \le \frac{i+1}{\delta}, \]
where the first inequality follows since each dense row contains strictly more than $\delta \frac{m_{j+1}}{m_j}$ columns out of the $m_{j+1}$ columns of~$\cC'_{m_{j+1}}$, and the last inequality is because $i+1 \ge m_j$ in the $j$th phase by definition.
By the same argument as before, we get the non-strict inequality $D \le \frac{i}{\delta}$ since $\delta$ is an integer.

The number of adjacent pairs of rows that are both non-tall and non-dense is at least $\lfloor (i+1)/2 \rfloor - T - D \ge i/2 - T - D$.
Since the algorithm halted unsuccessfully, each of these adjacent pairs must together contain strictly more than $d$ non-empty cells.
Therefore, the number of non-empty cells in~$M(\cR'_{i+1}, \cC'_{i+1})$ is at least
\[\left(\frac{i}{2} - \frac{i}{\delta} - \frac{i}{\delta}\right) \cdot d \ge \left(\frac{1}{2} - \frac{1}{20} - \frac{1}{20}\right) \cdot i \cdot d \ge \frac{2}{5} \cdot d \cdot \frac{i+1}{2} = c_\pi \cdot (i+1). \]

We obtain that $M(\cR'_{i+1}, \cC'_{i+1})$ contains $\pi$ by Lemma~\ref{lem:mt} and thus also $M$ contains~$\pi$, a contradiction.

\paragraph{Without knowledge of~$c_\pi$.}
It turns out that the algorithm is easily adapted to the case when $c_\pi$ is not known in advance.
In this case, we initially set $d = 1$ and double this parameter whenever there is no pair of adjacent rows or columns available for merging.
It is clear that the algorithm always generates a sequence of divisions $(\cR'_n, \cC'_n), \dots, (\cR'_1, \cC'_1)$, and it remains to observe that they have the desired properties.

\begin{figure}
	\textbf{Input:} $\pi$-avoiding permutation $\sigma$,  parameters $1 < m_1 < m_2 < \dots<  m_\ell < n$
	\begin{algorithmic}
		\State $d \gets 1, j \gets \ell$
		\State $(\cR, \cC) \gets$ the trivial division of~$M_\sigma$ into $n$ rows and columns of size~$1$
		\For{$i \gets n-1, \dots, 1$}
		\If{$Q_1$ or $Q_2$ is empty} \Comment{Triggers immediately in the first iteration.}
			\State $d \gets 2d$
			\ForAll{$T \in \cR \cup \cC$} 
			\Call{QueueForMerging}{$T$}
			\EndFor
		\EndIf
		\State $\code{MergeRows}(Q_1.\code{pop}())$ \Comment{Merge a pair of adjacent rows from~$Q_1$.}
		\State $\code{MergeCols}(Q_2.\code{pop}())$ \Comment{Merge a pair of adjacent columns from~$Q_2$.}
		\ForAll{$T \in B[i]$} \Comment{Queue non-tall rows, non-wide columns for merging.}
		\State{\Call{QueueForMerging}{$T.\code{prev}$}}
		\State{\Call{QueueForMerging}{$T$}} 
		\EndFor
		\If{$i = m_j$}
		\State $(\cR_j, \cC_j) \gets (\cR, \cC)$ \Comment{Copy current division.}
		\ForAll{$T \in \cR \cup \cC$} $T.\code{density} \gets 1$ \Comment{Reset density.}
		\EndFor
		\ForAll{$T \in \cR \cup \cC$} 
			\Call{QueueForMerging}{$T$}  \Comment{Add to queue non-dense.}
		\EndFor
		\State $j \gets j-1$
		\EndIf
		\EndFor
		
		\Procedure{QueueForMerging}{$T$}
		\LComment{Checks if $T$ can be merged with~$T.\code{next}$ and if yes, adds them to $Q_1$ or~$Q_2$.}
			\If{neither of $T, T.\code{next}$ is wide/tall, dense or in the queue}
			\If{$T.\code{cellsWithNext} \le d$}
			\State{$T.\code{inQueue} \gets \mathtt{true}, T.\code{next}.\code{inQueue} \gets \mathtt{true}$}
			\If{$T$ is a row} $Q_1.\code{push(T, T.\code{next})}$ \Comment{Push to the column queue.}
			\Else{} $Q_2.\code{push(T, T.\code{next})}$ \Comment{Push to the row queue.}
			\EndIf
			\EndIf
			\EndIf
		\EndProcedure
	\end{algorithmic}
	\caption{Algorithm for computing the structural decomposition.}\label{alg:decomposition}
\end{figure}

The properties \ref{item:rici:coarsening}--\ref{item:rici:non-dense} are independent of the value of~$d$ and thus, hold for $(\cR'_n, \cC'_n), \dots, (\cR'_1, \cC'_1)$ regardless of the value of~$d$.
Property~\ref{item:rici:sparse} implies that each row and column in every division contains at most $d_{\max}$ non-empty cells where $d_{\max}$ is the maximum (final) value of~$d$ throughout the algorithm.
We claim that $d_{\max} \le 10 c_\pi$.
Observe that otherwise the algorithm was unable to generate the sequence of divisions for $d = d_{\max}/2 \ge 5 c_\pi$ and we reach a contradiction exactly as in the case when $c_\pi$ was known.

\paragraph{Efficient implementation.}
We now proceed to describe an efficient implementation of the algorithm.
The current division~$(\cR'_i, \cC'_i)$ is stored in a structure where each row, column and non-empty cell is stored in a separate record that are interlinked with pointers.
The operation of merging a pair of adjacent rows or columns can then be implemented locally in time proportional to the number of non-empty cells involved.
We postpone the full details of this structure for now and focus first on the high-level overview.

Central to our implementation is the ability to detect mergeable pairs of rows and columns.
To that end, the algorithm maintains two queues $Q_1$ and $Q_2$ that store pointers to mergeable pairs of rows and columns, respectively.
It is important to observe that if two adjacent rows (or columns) can be merged within the division~$(\cR'_i, \cC'_i)$ then they can also be merged within any later division $(\cR'_j, \cC'_j)$ for $j >i$, assuming they still exist.
The algorithm will maintain the following two invariants pertaining to the queues~$Q_1$ and $Q_2$:
\begin{enumerate}[(Q1)]
\item each row appears at most once in~$Q_1$ and each column appears at most once in~$Q_2$, and\label[ipart]{item:queues:once}
\item for any pair of mergeable rows $R_1$ and $R_2$, at least one of $R_1$, $R_2$ is contained in~$Q_1$, and analogously for mergeable columns and $Q_2$.\label[ipart]{item:queues:all}
\end{enumerate}
As an immediate corollary, we get that $Q_1$ is empty if and only if there is no mergeable pair of rows and analogously for~$Q_2$ and columns.
Therefore in each step, the algorithm checks whether both~$Q_1$ and $Q_2$ are non-empty.
If yes, it pops a pair of mergeable rows from $Q_1$ and a pair of mergeable columns from~$Q_2$ and performs the respective merge operation.
Otherwise, it doubles the parameter~$d$ and resets~$Q_1$ and $Q_2$ by iterating over the whole current division.

Let us now describe how the algorithm maintains the invariants~\ref{item:queues:once} and~\ref{item:queues:all}.
Invariant~\ref{item:queues:once} is handled easily by storing a flag in each row and column record whether it is currently queued for merging.
Towards maintaining invariant~\ref{item:queues:all}, observe that a pair of adjacent rows~$R_1$ and $R_2$ can suddenly become mergeable within the division~$(\cR'_i, \cC'_i)$ in five different ways (the situation for columns is analogous):
\begin{enumerate}[(T1)]
\item one of $R_1$, $R_2$ was obtained by merging two rows in the previous step,\label[ipart]{item:rows:merging}
\item at least one of $R_1$, $R_2$ contains a non-empty cell in a column that was obtained by merging two columns in the previous step\label[ipart]{item:rows:orthogonal}
\item at least one of $R_1$, $R_2$ was tall with respect to~$(\cR'_{i+1}, \cC'_{i+1})$ and is no longer tall with respect to~$(\cR'_i, \cC'_i)$, and \label[ipart]{item:rows:tall}
\item at least one of $R_1$, $R_2$ was dense in the previous step but $i = m_j$ for some $j \in [\ell]$ and the algorithm enters next phase. \label[ipart]{item:rows:dense}
\item the parameter~$d$ was doubled at the beginning of the current step. \label[ipart]{item:rows:doubling}
\end{enumerate}

The easiest cases to handle are~\ref{item:rows:dense} and~\ref{item:rows:doubling} because the algorithm recomputes the queues~$Q_1$ and $Q_2$ from scratch whenever it doubles~$d$ or enters the next phase.
In the cases \ref{item:rows:merging} and~\ref{item:rows:orthogonal}, one of~$R_1$, $R_2$ must contain a non-empty cell that was involved in the last merge.
This will be detected and handled locally when merging rows or columns.
The only case that requires extra care to handle globally is case~\ref{item:rows:tall}.

A tall row~$R$ can appear only by merging two non-tall rows.
Observe that $R$ ceases to be tall in a future division $(\cR'_i, \cC'_i)$ where $i$ is the largest integer for which $|R| \le \frac{\delta n}{i}$ holds.
Therefore, we can immediately compute this index~$i$ upon the creation of~$R$ as $i = \lfloor \frac{\delta n}{|R|} \rfloor$.
The algorithm stores an extra array~$B$ of length~$n$ where $B[i]$ is a list that collects precisely all rows (resp.~columns) that are tall (resp.~wide) within $(\cR'_{i+1}, \cC'_{i+1})$ but no longer within~$(\cR'_i, \cC'_i)$.
So whenever a tall row~$R$ (resp.~wide column) is obtained by merging, it is added to the list $B[i]$ for $i = \lfloor \frac{\delta n}{|R|} \rfloor$.
Then in the $j$-th step, the algorithm first iterates over the list $B[j]$, checking whether some of the rows (resp. columns) therein can be merged with their neighbors and adding all such pairs to the queue~$Q_1$ (resp.~$Q_2$).
Observe that this does not incur any additional overhead to the overall runtime since we detect each tall row (resp.~wide column) and store it in $\fO(1)$ time upon its creation and it is then once removed from the array~$B$, again at constant cost.

This concludes the global overview of the implementation.
Now, we proceed to specify the data stored for each row, column and non-empty cell which then allow an implementation of $\code{MergeRows}$ and $\code{MergeCols}$ methods in linear time with respect to the non-empty cells involved in the operation.

\paragraph{Row/column structure.}
The structure for columns and rows is analogous and we describe its semantics only for rows. 
Formally, each row~$R$ is represented by a record \code{Strip} consisting of the following:
\begin{itemize}
\item \code{prev} and \code{next}: Pointers to the previous and next row in the top-to-bottom order (or \texttt{null} if this is the first or the last row). Thus, all rows form a doubly-linked list.
\item \code{id}: Unique identifier of the row from the set $[n]$ that respects the top-to-bottom order of rows, i.e., we have $R.\code{id} < R.\code{next}.\code{id}$ whenever $R.\code{next} \neq \mathtt{null}$.
\item \code{cells}: Number of non-empty cells in~$R$.
\item \code{firstCell}: Pointer to the record of the leftmost non-empty cell in the row.
\item \code{size}: The height of the row $|R|$, i.e., the number of original rows from~$M$ contained in~$R$.
\item \code{inQueue}: Boolean flag whether this row is already queued for merging. That is $R.\code{inQueue} = \mathtt{true}$ if $R$ appears in $Q_1$ and $\mathtt{false}$ otherwise.
\item \code{density}: Number of rows of the previous division~$(\cR'_{m_j}, \cC'_{m_j})$ contained in~$R$.
\item \code{cellsWithNext}: Number of non-empty cells in the union of $R$ with $R.\code{next}$ or $\infty$ if $R$ is the last row ($R.\code{next} = \mathtt{null}$).
\end{itemize}

\paragraph{Cell structure.}
Each non-empty cell~$C$ is represented by a record \code{Cell} consisting of the following:
\begin{itemize}
	\item \code{left} and \code{right}: Pointers to the previous and next non-empty cell in the same row, or $\mathtt{null}$ if there is no such cell.
	\item \code{below} and \code{above}: Pointers to the previous and next non-empty cell in the same column, or $\mathtt{null}$ if there is no such cell.
	\item \code{row} and \code{column}: Pointers to the row and column containing the cell~$C$.
\end{itemize}

\paragraph{Merging operation.}
We describe the implementation of $\code{MergeRows}(R_1, R_2)$ in linear time with respect to the number of non-empty cells in $R_1$ and $R_2$, the merging of columns is analogous.
First, we initialize a record \code{Strip} for a new row~$R$ and replace $R_1$ and $R_2$ with~$R$ by setting the fields \code{prev} and \code{next} appropriately at~$R$ and its at most two adjacent rows.
Second, we set $R.\code{id}$ to $R_1.\code{id}$, $R.\code{size}$ to $R_1.\code{size} + R_2.\code{size}$ and $R.\code{density}$ to $R_1.\code{density} + R_2.\code{density}$.

Next, we simultaneously traverse in left-to-right order the non-empty cells in both rows~$R_1$ and $R_2$ that can be accessed through $R_1.\code{firstCell}$ and $R_2.\code{firstCell}$.
Let us first describe the traversal ignoring the necessary updates to $\code{cellsWithNext}$ fields.
Crucially, for a pair of cells~$C_1$ in~$R_1$ and $C_2$ in~$R_2$ we can determine in constant time whether they share the same column or which one is more to the left by comparing $C_1.\code{column}.\code{id}$ with $C_2.\code{column}.\code{id}$.
This allows the procedure to simultaneously traverse cells in both rows in the left-to-right order while relinking pointers or replacing two non-empty cells in the same column with a single new non-empty cell.
Along the way, it counts the number of non-empty cells in~$R$ and sets $R.\code{cells}$ appropriately afterwards.

However, we still need to handle updating $\code{cellsWithNext}$ fields in rows $R$, $R.\code{prev}$ (if it exists) and potentially every column such that the cell in its intersection with~$R$ is non-empty (cf.\ \ref{item:rows:merging} and \ref{item:rows:orthogonal}).
To arrange the former, the procedure traverses $R$ at most two more times at the end, once per each adjacent row to update $\code{cellsWithNext}$ fields and potentially add $R$ with either of its neighbors to~$Q_1$ for merging.
For the latter, we modify the merging traversal of $R_1$ and $R_2$ to always consider 2 non-empty cells from both $R_1$ and $R_2$.
Out of these (at most) four cells, it is decidable in constant time how $\code{cellsWithNext}$ changes for the affected columns and in extension, whether they should be added to~$Q_2$ for merging.

\paragraph{Running time.}
Finally, let us bound the running time.
First, the algorithm must construct the initial division $(\cR'_n, \cC'_n)$ out of the $\pi$-avoiding permutation $\sigma$ on input.
This can be done straightforwardly in linear time using access to $\sigma$ in sorted orders by both indices and values.
This takes $\fO(n \lg c_\pi)$ time using the optimal algorithm for sorting pattern-avoiding sequences~\cite{Opler_sort}.

In the following, let $d_i = 2^i$ denote the value of $d$ after doubling $i$ times and let $i_{\max}$ be the index such that the algorithm finished with the value of $d$ equal to $d_{i_{\max}}$.
Recall that we have $i_{\max} \in \fO(\lg c_\pi)$ since $d_{i_{\max}} \le 10 c_\pi$.
We denote by $n_1, \dots, n_{i_{\max}}$ the sizes of divisions where the algorithm doubled the parameter~$d$.

We claim that $n_i \in \fO(n/d_i)$.
To show this, focus at the step when the doubling $d_{i-1} \to d_i$ occurred.
First, observe that we have $n_i \ge 2$ since the algorithm have not finished at that point yet.
By our previous arguments, there are at most $n_i/\delta$ tall rows and at most $n_i/\delta$ dense rows.
This leaves at least $\lfloor\frac{n_i}{2}\rfloor- \frac{2}{\delta} n_i \ge \frac{3}{20} n_i$ pairs of adjacent rows that are neither tall nor dense.
Since none of these pairs can be merged, each such pair contains together strictly more than $d_{i-1}$ non-empty cells.
Therefore, there are at least $d_{i-1} \cdot \frac{3}{20} n_i \ge \frac{1}{20} d_i n_i$ non-empty cells in total and since the number of non-empty cells is trivially at most~$n$, we obtain $n_i \le \frac{n}{20 d_i} \in \fO(n/d_i)$.

Let us now bound all the time incurred during every doubling of the parameter $d$.
Every doubling of~$d$ causes an iteration over all rows and columns of the current division and adding mergeable pairs to the respective queues.
Therefore, this takes time linear in the number of rows and columns which makes
$\fO(\frac{n}{d_1} + \dots + \frac{n}{d_{i_{\max}}}) \subseteq \fO(n)$ time in total.

Next, we bound the total time taken by all calls to $\code{MergeRow}$ and $\code{MergeColumn}$.
Recall that a single call takes linear time with respect to the number of non-empty cells participating in the merge.
We fix $i \in [i_{\max}]$ and derive an upper bound on all such calls that occurred when $d$ was equal to $d_i = 2^i$.
After the doubling step $d_{i-1} \to d_i$, there were at most $\fO(n/d_i)$ rows and columns.
The algorithm only merges rows and columns if their union contains at most $d_i$ non-empty cells which makes a single call to take $\fO(d_i)$ time. 
Moreover, each merge decreases either the number of rows or columns by one and thus, there are at most $\fO(n/d_i)$ merges with combined runtime of $\fO(n)$.
Together, this bounds the total time of all merging to $\fO((i_{\max} + 1) \cdot n) = \fO((\lg c_\pi + 1) \cdot n)$.

Finally, it remains to bound the time needed to copy and output the divisions $(\cR_j, \cC_j)$ for every $j \in [\ell]$ and all the other extra work done at the end of each phase.
At the end of the $j$-th phase, the algorithm writes out the current division of size $\fO(c_\pi \cdot (|\cR_j| + |\cC_j|)) = \fO(c_\pi \cdot m_j)$ and iterates over all rows and columns in $\fO(m_j)$ time.
In total, this takes $\fO(c_\pi \cdot \sum_{j=1}^\ell m_j)$ time as promised.

\section{Proof of the treewidth structural lemma}\label{sec7}
In this section, we prove the balanced decomposition result for permutations of bounded treewidth (\Cref{lem:decomp_tw}).

\paragraph{Grid-width.}
In order to go from the treewidth of incidence graph~$G_\sigma$ to divisions, we exploit a functionally equivalent parameter \emph{grid-width}~\cite{gridwidth}.
For any subset~$S$ of 1-entries in $M_\sigma$, the \emph{grid complexity} of~$S$ is the smallest number~$d$ such that there are at most $d$ column intervals and at most $d$ row intervals such that $S$ lies in the intersection of these rows and columns while no other 1-entry lies in these columns or rows.
Informally, there exists a division $(\cR_, \cC)$ where $S$ is precisely a union of several components that together span at most $d$ rows and $d$ columns.
Observe that if $S$ and $T$ are sets of grid complexity at most~$d$, then their union $S \cup T$ and difference $S \setminus T$ both have grid complexity at most~$2d$.

A \emph{grid tree}~$T$ of~$\sigma$ is a rooted tree with $n$ leaves, each leaf being marked with a distinct entry $(i, \sigma_i)$ of~$\sigma$.
For a node $v$ in~$T$, let $S^T_v$ denote the entries of~$\tau$ that appear as leaf labels in the subtree of~$T$ rooted in~$v$.
The grid-width of~$T$, denoted by $\gw^T(\sigma)$  is the maximum grid complexity of $S^T_v$ over all nodes of~$T$.
And as usual, \emph{grid-width} of~$\sigma$, denoted by $\gw(\sigma)$, is the minimum $\gw^T(\sigma)$ over all grid trees of~$\sigma$.
Observe that we can without loss of generality assume that every inner node in~$T$ has exactly two children.
It is known that for every permutation~$\sigma$, we have $\frac{1}{8} \tw(G_\sigma) \le \gw(\sigma) \le \tw(G_\sigma) + 2$, where $\tw$ denotes treewidth. Moreover, there is a linear-time algorithm that receives a tree decomposition of width~$k$ for $G_\sigma$ on input and outputs a grid tree~$T$ such that $\gw^T(\sigma) \le k + 2$~\cite[Proposition 4.1]{gridwidth}.

\medskip

As a warm-up, let us argue that a grid tree~$T$ for~$\tau$ of small grid-width~$d$ can be easily turned into a hierarchy of divisions with small components if we ignore the balancedness constraints (\ref{item:bal-gridding-tw:width-rows} and \ref{item:bal-gridding-tw:width-density}).
We again start with the trivial division into single rows and columns of~$M_\tau$.
Our strategy is to contract edges adjacent to leaves in~$T$ one by one until we arrive at a singleton tree, and mirror these contractions in the divisions.
Along the way, each node~$v$ in the tree has its associated set of $\tau$-entries $S(v)$ such that (i) the grid complexity of $S(v)$ is at most~$3d$, and (ii) the sets $S(v)$ form a partition of~$\tau$.
At the beginning, we set $S(v)$ to be empty for the internal nodes of~$T$ and for a leaf~$v$, we set $S(v) = \{(i, \tau_i)\}$ where $(i, \tau_i)$ is the label of~$v$.
In every step, we choose a leaf~$v$ with parent~$w$ in the current tree, remove the leaf~$v$ and set $S(w) \gets S(v) \cup S(w)$.

Let us argue that the grid complexity of the sets~$S(v)$ remains bounded throughout the process.
To see that, notice that if we trace the contractions back, each node~$v$ corresponds to a connected subtree~$T_v$ of the original grid tree~$T$.
Moreover, this subtree $T_v$ can be obtained from~$T$ by taking a subtree rooted in some node~$v_1$ and then removing at most two of its rooted subtrees since $v$ has at most two children.
But then $S(v)$ is equal to $S^T_{v_1} \setminus (S^T_{v_2} \cup S^T_{v_3})$ where all $S^T_{v_1}, S^T_{v_2}, S^T_{v_3}$ are sets of grid complexity at most~$d$.
It follows that $S(v)$ has grid complexity at most $3d$.

Finally, we describe how we update the division at each contraction.
Throughout, we maintain the smallest division such that for every two different nodes $v$ and $w$, the sets $S(v)$ and $S(w)$ occupy pairwise different rows and columns.
Therefore, when we delete a leaf $v$ with a parent~$w$, we merge every adjacent pair of rows and columns such that one of them intersects $S(v)$ and the other $S(w)$.
This guarantees that every component eventually spans at most $3d$ rows and columns of the division because of our bound on the grid complexity of the sets~$S(v)$.
However, that does not directly bound the number of rows and columns occupied by a single component throughout the whole process.
That is because after contracting in~$T$, we start merging rows and columns one by one and thus, it might happen that the new component initially spans $2 \cdot 3d$ rows and columns before being pushed down to at most $3d$ through row and column merges.

We omit detailed description of an efficient implementation and include it only for the full, balanced, variant.

\medskip

From a very high-level view, we first compute a grid tree~$T$ of~$\tau$ with small grid-width that we then use to guide the creation of hierarchy of divisions.
In the general decomposition lemma, we performed merging operations on adjacent rows and columns greedily as long as they maintained given invariants.
Here, we use the grid tree~$T$ as a guide to which pairs of columns and rows can be merged at each step similarly to the warm-up.
However, we allow contraction of not only the edges incident to leaves but also edges incident to vertices with a single child.
This procedure follows the common approach for computing a balanced treewidth decomposition~\cite{BodlaenderH98} based on parallel tree contractions~\cite{MillerR89}.
In this way, there is still enough freedom in choosing the next contraction which allows us to maintain the balancedness constraints in the same vein as before.

The first step of the decomposition is to get a grid tree of low grid-width.
To do that, we invoke any single exponential approximation algorithm for treewidth -- say the 2-approximation due to Korhonen~\cite{Korhonen21} -- and immediately turn it into a grid tree with grid-width~$t'$ where $t' \le 2 \tw(G_\sigma) + 2$~\cite[Proposition 4.1]{gridwidth}.

Instead of computing all the divisions $(\cR_1, \cC_1), \dots, (\cR_\ell, \cC_\ell)$ individually, the algorithm again just dynamically iterates through a sequence of divisions $(\cR'_n, \cC'_n), \dots, (\cR'_1, \cC'_1)$ such that
\begin{enumerate}[(i)]
	\item The division $(\cR'_{i}, \cC'_{i})$ is a coarsening of $(\cR'_{i+1}, \cC'_{i+1})$.\label[ipart]{item:rici-tw:coarsening}
	\item The division $(\cR'_{i}, \cC'_{i})$ has exactly $i$ rows and columns.\label[ipart]{item:rici-tw:rows-cols}
	\item Each row (resp.\ column) of $M(\cR'_i,\cC'_i)$ has 
	height (resp.\ width) at most $96 t' \cdot n/i$.\label[ipart]{item:rici-tw:width-height}
	\item Each row and column in~$M(\cR'_i,\cC'_i)$ is obtained from merging at most $192 t' \cdot m_{j+1}/m_j$ rows or columns of~$M(\cR'_{m_{j+1}}, \cC'_{m_{j+1}})$,
	where $j$ is the largest integer such that $m_j \le i$.\label[ipart]{item:rici-tw:non-dense}
	\item Each component in~$(\cR'_i, \cC'_i)$ spans at most $12t + 12$ rows and columns.
	\label[ipart]{item:rici-tw:sparse}
\end{enumerate}
We get the desired output by copying out $(\cR_i, \cC_i) = (\cR'_{m_i}, \cC'_{m_i})$ for every $i \in [\ell]$.

We start  with $(\cR'_n, \cC'_n)$ being the trivial (finest) division where each row (resp.\ column) contains exactly one row (resp.\ column) of the input matrix~$M$.
As in the warm-up, we associate to each node $v$ in the tree~$T$ a set $S(v)$, initially set to consist of the labels in leaves  and empty elsewhere.
The division at hand is maintained as a smallest division with equal number of rows and columns that separates the sets $S(\cdot)$, i.e., there is no column or row that intersect $S(v), S(w)$ for two different vertices.
The run is again split into $\ell+1$ phases with the $j$-th phase generating divisions $(\cR'_i, \cC'_i)$ for $i \in \{m_j, \dots, m_{j+1}-1\}$.
However, the size and density conditions are now imposed on the sets~$S(\cdot)$ instead of rows and columns of the division.
A vertex~$v$ in~$T$ is \emph{large} if $|S(v)| > 8 \cdot n/ |V(T)|$ where $V(T)$ is the set of vertices in~$T$.
Moreover, $v$ is \emph{dense} in the $j$-th phase if $S(v)$ is a union of more than $192 t' \cdot m_{j+1}/m_j$ rows or columns of the division $(\cR'_{m_{j+1}}, \cC'_{m_{j+1}})$.

We describe row and column merges in batches which are performed in immediate succession.
However, if a division $(\cR'_i, \cC'_i)$ of size exactly $i = m_j$ for some~$j \in [\ell]$ is encountered, the algorithm outputs the current division and resets density of vertices in~$T$ before continuing with individual row and column merges.
In one step, the algorithm finds a vertex $v$ with parent $w$ such that $v$ is either a leaf or has a single child and, moreover, $v$ and $w$ are both neither large nor dense.
We contract the edge $\{v,w\}$ in~$T$ and set $S(v') \gets S(v) \cup S(w)$ where $v'$ is the vertex obtained in the contraction.
Afterwards, the algorithm keeps merging adjacent rows and columns that contain~$S(v')$ where crucially, it alternates between merging rows and columns and stops when another row or column merge is no longer possible.
The reason for this is to maintain the equality between number of rows and columns in the divisions.

This process continues all the way until a trivial division~$(\cR'_1, \cC'_1)$ is reached.

\paragraph{Correctness.}
We have to verify that the algorithm is guaranteed to finish, i.e., there will always be a suitable pair of vertices $v,w$ in~$T$ to merge, and properties \ref{item:rici-tw:coarsening}--\ref{item:rici-tw:sparse} hold for every division $(\cR'_i, \cC'_i)$.


First, we show that \ref{item:rici-tw:coarsening}--\ref{item:rici-tw:sparse} hold for every division.
Properties \ref{item:rici-tw:coarsening} and \ref{item:rici-tw:rows-cols} are maintained directly by the algorithm.
Towards showing \ref{item:rici-tw:sparse}, recall that for a vertex~$v$ in the tree~$T$ we denote by~$T_v$ the connected subtree of the original grid tree that was contracted into~$v$.
We maintain that every vertex in~$T$ has at most two children and thus, $T_v$ is obtained by taking a rooted subtree and then removing up to two of its rooted subtrees.
Therefore, $S(v)$ is equal to the set $S^T_{v_1}$, $S^T_{v_1} \setminus S^T_{v_2}$, or $S^T_{v_1} \setminus (S^T_{v_2} \cup S^T_{v_3})$ for some $v_1, v_2, v_3$ in the original grid tree and the grid complexity of $S(v)$ is at most $3t' \le 6t + 6$.
As in the warm-up, this implies that in the worst case a single component can contain at most $2 \cdot 3t' = 12t + 12$ rows and columns following some edge contraction before being pushed down to at most $3t' = 6t + 6$ through row and column merges.

Properties \ref{item:rici-tw:width-height} and \ref{item:rici:non-dense} follow similarly to their analogues in \Cref{lem:decomp}.
Specifically, we have $|S(v)| \le 16 n/ |V(T)|$ for every vertex~$v$ in~$T$ since we never contract vertices with $S(v) > 8 n/ |V(T)|$ and $|V(T)|$ only decreases throughout the process.
We obtain $|S(v)| \le 96 t' n/ i$ by plugging in the inequality $|V(T)| \ge i/(6t')$ that is due to~\ref{item:rici-tw:sparse}.
For analogous reasons, any component in the $j$-th phase occupies at most $192 \cdot m_{j+1}/m_j$ rows and columns of the division $(\cR'_{m_{j+1}}, \cC'_{m_{j+1}})$ and the same bound naturally transfers to every individual row or column in the component.

Finally, we show that the tree~$T$ always contains a suitable pair of vertices to merge.
Assume for a contradiction that there was no suitable pair of vertices in~$T$ to merge at some step and let $n'$ denote the number of vertices in~$T$ at that moment.
As we already argued, there are at least $n'/2$ vertices with at most one child.
Each such vertex is a candidate for merging together with its parent.
Moreover, each vertex participates in at most two such mergeable pairs of adjacent vertices -- this happens when both children of a vertex are leaves, or there is vertex that has a single child which is itself a degree-2 vertex or a leaf.
Thus, there are at least $n'/4$ disjoint pairs of vertices $v,w$ in~$T$ such that $w$ is a parent of~$v$ and $v$ is either a leaf or has a single child.

Let us denote by $L$ and $D$ the number of large vertices and dense vertices, respectively.
We have $L < n'/8$ since $|S(v)| > 8 n/n'$ for every large vertex~$v$.
The number of dense vertices can be bounded as
\[ D < \frac{2 m_{j+1}}{96 t'\, \frac{m_{j+1}}{m_j}} \le \frac{m_{j+1}}{48 t'\, \frac{m_{j+1}}{m_j}} \le \frac{m_j}{48 t'} \le \frac{n'}{8}\]
where the first inequality holds since each dense component contains strictly more than $96 t'\, \frac{m_{j+1}}{m_j}$ rows or columns of $(\cR'_{m_{j+1}}, \cC'_{m_{j+1}})$ and the third inequality holds since the number of rows and columns in the current division is at least $m_j$ and at most $6t'n'$ (and, thus, $m_j \le 6t'n'$).
As a result, there are strictly less than $n'/4$ large and dense vertices in~$T$ so there is always at least one possible edge that can be contracted.

\paragraph{Implementation.}
The algorithm can be implemented similarly to the general decomposition lemma with the advantage that we can afford more dependence on the treewidth~$t$ in the runtime.
This is because the initial approximation of treewidth already takes $2^{\fO(t)} n$ time and thus, we only aim at a runtime $\fO_t(n)$.
Therefore, we do not describe the implementation in full detail and mostly refer to the implementation in \Cref{lem:decomp}.

Recall that the algorithm in \Cref{lem:decomp} maintains two queues $Q_1$ and $Q_2$ that store adjacent pairs of rows and columns for future merging.
Since we contract edges in the tree~$T$, we now maintain a single queue of edges~$Q$ available for contraction together with size, density information stored in each vertex of~$T$.
This allows us to spend only $\fO(1)$ time to pop the next edge for contraction, update~$T$ and potentially add the new vertex with some of its (at most three) neighbors to~$Q$.
As before, we maintain a bucket structure to keep track of large vertices and reintroduce them back when they cease to be large, without additional overhead.
For the density, we allow a complete iteration over the tree at the end of each phase which resets the density of each dense vertex.

We store the division in the same data structure, only omitting \code{size}, \code{inQueue}, \code{density} and \code{cellsWithNext} fields in every row and column.
Additionally, each vertex~$v$ in~$T$ stores pointers to every row and column that is intersected by~$S(v)$.
After contracting an edge~$\{v,w\}$ in~$T$, the algorithm iterates through the rows and columns containing $S(v), S(w)$ and merges them one by one alternating between rows and columns.
This can be straightforwardly implemented in $\fO(t^2)$ time where one factor comes from the number of rows and columns occupied by each component and the other from merging individual adjacent pairs of rows and columns.
Overall, there are $\fO(n)$ edge contractions each incurring $\fO_t(1)$ time for an overall runtime of $\fO_t(n)$.


\small
\bibliographystyle{alphaurl}
\bibliography{main}
\end{document}